\documentclass{emulateapj}
\usepackage{graphicx, amsmath, amssymb, amsthm}

\makeatother

\newcommand\etal{{\it et~al.~}}
\newcommand\kms{ {\rm km\, s$^{-1}$ } }

\begin{document}
\title{A Multistep Algorithm for the Radiation Hydrodynamical Transport of Cosmological Ionization Fronts
and Ionized Flows}
%%\author{Daniel Whalen}
%%\affil{Department of Physics, University of Illinois, Urbana-Champaign, Urbana, IL  61801}
%%\email{dwhalen@cosmos.ucsd.edu}
%%\author{Michael L. Norman}
%%\affil{University of California, San Diego, La Jolla, CA 92093}

\author{Daniel Whalen\altaffilmark{1,2,3} \& Michael L. Norman\altaffilmark{1}}
\altaffiltext{1}{Center for Astrophysics and Space Sciences,
University of California at San Diego, La Jolla, CA 92093, U.S.A.
Email: dwhalen@cosmos.ucsd.edu}
\altaffiltext{2}{Department of Physics, University of Illinois in
Urbana-Champaign}
\altaffiltext{3}{Theoretical Astrophysics (T-6), Los Alamos National
Laboratory}

\begin{abstract} 

Radiation hydrodynamical transport of ionization fronts in the next generation of cosmological 
reionization simulations holds the promise of predicting UV escape fractions from first principles 
as well as investigating the role of photoionization in feedback processes and structure formation.  We 
present a multistep integration scheme for radiative transfer and hydrodynamics for accurate 
propagation of I-fronts and ionized flows from a point source in cosmological simulations.  The 
algorithm is a photon-conserving method which correctly tracks the position of I-fronts at much 
lower resolutions than non-conservative techniques.  The method applies direct hierarchical updates 
to the ionic species, bypassing the need for the costly matrix solutions required by implicit methods 
while retaining sufficient accuracy to capture the true evolution of the fronts.  We review the physics 
of ionization fronts in power-law density gradients, whose analytical solutions provide excellent 
validation tests for radiation coupling schemes.  The advantages and potential drawbacks of direct and 
implicit schemes are also considered, with particular focus on problem timestepping which if not 
properly implemented can lead to morphologically plausible I-front behavior that nonetheless departs 
from theory.  We also examine the effect of radiation pressure from very luminous central sources on the 
evolution of I-fronts and flows. 

\end{abstract}

\keywords{cosmology: theory---early universe---H II regions: simulation}

\section{Early Reionization}

The 2003 WMAP discovery of large optical depths to electron scattering at z $\sim$ 15 \citep{ket03,
spet} motivated several numerical and analytical studies of early reionization \citep{hh03,c03,sl03,
soet03,cfw03,cfmr01,wl03}.  The goal of these models was to extend the classical picture of
reionization to account for the large electron fractions at high redshifts while still 
accommodating lower redshift observations.  In particular, Sokasian, \etal and Ciardi, \etal 
employed large scale cosmological simulations combining gas and dark matter dynamics with radiative 
transfer to follow the growth of ionized regions in the early IGM.  Global electron densities in 
these calculations were integrated over redshift to compute the Thomson scattering optical depth 
$\tau_e$ for a variety of scenarios.

\subsection{Limits to Current Models}

All these simulations assume several free parameters: UV escape fractions, Pop III stellar 
masses, the positive and negative feedback of one generation of UV sources upon the next, and 
the Pop III/Pop II rollover in mass spectrum and UV production with redshift.  Furthermore, 
there was no hydrodynamic response to the energy deposited into the gas by the passage of fronts 
in these models, seriously altering the evolution of the fronts themselves.  The studies cited 
above postprocessed successive hydrodynamic snapshots of the IGM with radiative transfer without 
energy deposition into the gas to evolve primordial H II regions (see \citet{mfc03} for a summary 
of the latest algorithms applied to cosmological RT).  The challenge of the next generation of
early reionization simulations is to capture the radiation hydrodynamics of ionization and feedback
physics on small scales to determine the final sizes and distribution of I-fronts in the large
simulation volumes necessary for statistically accurate structure formation \citep{bl04}.

Upcoming observations able to discriminate between early reionization scenarios underscore 
the need for ab initio simulation of the early IGM, in which reionization properly unfolds over many 
redshifts and generations of luminous objects.  21 cm line observations in both emission and absorption 
(by the Square Kilometer Array) could yield cosmic electron fraction profiles as a function of redshift,  
beyond current WMAP and upcoming Planck measurements that are limited to electron column densities.  If 
foreground contamination can be overcome, these observations might also unveil the size and morphologies 
of early H II regions \citep{fb04}.  Signatures of Pop III stars manifest as excesses in the near-IR 
cosmic background may soon be measured by balloon and satellite missions \citep{sbk02,cet04,cy04}.  JWST 
will also open the first direct observational window on protogalaxies with a few thousand Pop III stars 
at 15 $<$ z $<$ 20.

\subsection{Motivation for Cosmological Radiation Hydrodynamics}

The escape of ionizing UV photons from primordial minihalos and protogalaxies is mediated by the 
hydrodynamical transitions of ionization fronts on sub-parsec scales \citep{wan04,ket04}, and failure
to properly capture their breakout can alter the final extent of H II regions on kiloparsec scales.  In 
some cases static transfer completely fails to predict the exit of fronts from early galaxies by 
excluding the gas motions that can free them \citep{wl00}.  Radiation hydrodynamical simulations can predict escape 
fractions in the next generation of models by following I-fronts as they begin deep within primordial 
structures and blossom outward to ionize the IGM, accurately resolving their true final sizes and 
morphologies. 

Coupled to reactive networks able to evolve primordial H$_2$ chemistry, radiation hydrodynamics will 
also better model the radiative feedback mechanisms known to operate in the early universe, which 
remain to be incorporated in detail in large scale calculations.  Local entropy injection by UV sources 
\citep{oh03}, Lyman-Werner dissociation of H$_{2}$ in minihalos and protogalaxies, and catalysis of 
molecular hydrogen by free electrons \citep{rgs01,oet05,mba03} are key processes governing the rise 
of early star populations and the high-redshift ionizing background.  On small scales radiation 
hydrodynamics will also resolve ionized gas outflows in H II regions that facilitate the dispersal of 
metals from the first supernovae, exhibit dynamical instabilities potentially leading to clumping and 
further star formation, and limit the growth of black holes left in minihalos.  Resolving 
radiative feedback over a few generations of primordial stars will enable their inclusion in large 
simulations over many redshifts with confidence later on.
 
On large scales radiation hydrodynamics will be crucial to determine whether IGM photoheating cascades 
from small to large scales through nonlinear dynamical evolution to affect structure formation at later 
redshifts.  The expansion of cosmological ionization fronts through filaments and voids is also inherently
hydrodynamical in nature, as is the photoevaporation of minihalos that can impede these fronts \citep{sir04}.  
Static transfer cannot reproduce the outflows confirmed by numerical simulations to enhance the 
photoionization of these structures and therefore understimates the advance of I-fronts into the early IGM.  

\subsection{Overview}

To investigate the numerical issues confronting the incorporation of radiation hydrodynamics into future 
large scale structure evolution models we have developed an explicit multistep scheme for ionization front 
transport from a single point source in the ZEUS-MP hydrocode \citep{n00}.  The issues fall into two categories.  The 
first is how to calculate photoionization rates everywhere on the numerical grid, whether by ray tracing, 
variable tensor Eddington factors, flux limited diffusion, or Monte Carlo approaches for either single or 
multiple sources (see \citet{rc05,ret05} for novel raytracing schemes for the Enzo and FLASH adaptive mesh 
refinement (AMR) codes).  In section 2 we examine the second issue:  how to couple reaction networks 
and energy equations driven by ionization to the hydrodynamics, which has only recently begun to be examined 
by the cosmology community \citep{wwd05}.  Our algorithm easily extends to 
multiple frequencies and can be readily interfaced with transfer techniques accommodating many point sources.  
In sections 4 and 5 we present a comprehensive suite of static and hydrodynamic I-front test problems utilized to benchmark 
our code that can be applied to validate future methods.  In particular, the hydrodynamic benchmarks are 
adopted from an analytical study of ionization fronts in power-law density profiles done by \citet{ftb90} (see 
\citet{y86} for a thorough review of numerical and analytical studies of classical H II regions).  The tests 
encompass the range of I-front dynamics likely to occur in cosmological settings and will challenge the 
versatility and robustness of any code.  They also expose many features of ionized flows that are exhibited
by any density gradient.

We examine in section 6 how individual zones approach ionization equilibrium as well as the timescales that govern each 
phase of I-front and ionized flow evolution in a variety of density regimes.  
We discuss how these timescales control the timestep advance of the numerical solution and explore avenues
for future algorithm optimization.  The impact of radiation pressure on I-fronts and flows is also reviewed in section 7, 
and we provide an array of improved UV escape fraction calculations for Pop III stars in section 8.

\section{Numerical Algorithm}

Our modified ZEUS-MP hydrocode solves explicit finite-difference approximations to Euler's equations of fluid 
dynamics together with a 9-species primordial gas reaction network \citep{anet97} that utilizes photoionization rate coefficients computed 
by a ray-casting radiative transfer module.  Ionization fronts thus arise as an emergent feature of reactive 
flows and radiative transfer in our simulations and are not tracked by computing equilibria positions along 
lines of sight, as is done in many algorithms.  The fluid dynamics equations are
\vspace{0.1in}
\begin{align}
\frac{\partial \rho}{\partial t}  & =  - \nabla \: \cdotp \; (\rho {\bf v}) \notag \\
\frac{\partial \rho v_{i}}{\partial t}  & =  - \nabla \: \cdotp \; (\rho v_{i} 
{\bf v}) \: - \: \nabla p \: - \: \rho \nabla \Phi \: - \: \nabla \cdotp {\bf Q}    \\ 
\frac{\partial e}{\partial t}  & =  - \nabla \: \cdotp \; (e {\bf v}) \: - \: p\nabla \: 
\cdotp \: {\bf v} \: - \: {\bf Q} : \nabla  {\bf v} \: + \: {\dot{e}}_{rad} \notag 
\end{align} \vspace{0.05in} 

\noindent 
where $\rho$, e, and the v$_{i}$ are the mass density, internal energy density, and velocity of each zone and 
p (= ($\gamma$-1) e) and {\bf{Q}} are the gas pressure and the von Neumann-Richtmeyer artificial viscosity 
tensor \citep{ns92}.  ${\dot{e}}_{rad}$ represents radiative heating and cooling terms described below.  The 
left-hand side of each equation is updated term by term in operator-split and directionally-split 
substeps, with a given substep incorporating the partial update from the previous substep.  The gradient (force) 
terms are computed in the ZEUS-MP source routines and the divergence terms are calculated in the ZEUS-MP 
advection routines \citep{ns92}.

\subsection{Reaction Network}

The primordial species added to ZEUS-MP (H, H$^{+}$, He, He$^{+}$, He$^{2+}$, H$^{-}$, 
H$^{+}_{2}$, H$_{2}$, and e$^{-}$) are assumed to share a common velocity field 
and are evolved by nine additional continuity equations and the nonequilibrium rate equations 
of \citet{anet97} \vspace{0.05in}
\begin{equation}
\frac{\partial \rho_{i}}{\partial t} = - \nabla \: \cdotp \; (\rho_{i} {\bf v}) 
+ \sum_{j}\sum_{k} {\beta}_{jk}(T){\rho}_{j}{\rho}_{k} - \sum_{j} {\kappa}
_{j}{\rho}_{j} \vspace{0.05in}
\end{equation}
where ${\beta}_{jk}$ is the rate coefficient of the reaction between species j and k that creates (+) or 
removes (-) species i, and the ${\kappa}_{j}$ are the ionization rates.  The divergence term for each 
species is evaluated in the advection routines, while the other terms form a reaction network which is 
solved separately from the source and advection updates.  To focus our discussion on the gas dynamics of 
I-fronts our present calculations take the primordial gas to be hydrogen only, which can be ionic or 
neutral but not molecular.  The rate equations reduce to \vspace{0.05in}
\begin{equation}
\displaystyle\frac{dn_H}{dt} = k_{rec}n_{H^+} n_{e} - k_{ph}n_{H} - k_{coll}n_
{H}n_{e} \vspace{0.10in}  
\end{equation} 
where k$_{rec}$, k$_{ph}$, and k$_{coll}$, are the rate coefficients for recombination, photoionization 
and electron collisional ionization of hydrogen and the n$_{i}$ are the number densities. 

When the full reaction network is activated we enforce charge and baryon conservation at the end of each hydrodynamic cycle with the following 
constraints \vspace{0.1in}
\begin{align}
{\rho}_{b}f_{H} & = {\rho}_{H} + {\rho}_{H^{+}} + {\rho}_{H^{-}} + {\rho}_{{H_{2}}^{+}} + {\rho}_{H_{2}}   \\
m_{H}n_{e} & = {\rho}_{H^{+}} - {\rho}_{H^{-}} + {\frac{1}{2}}{\rho}_{{H_{2}}^{+}} + {\frac{1}{4}}{\rho}_{He^{+}}+ 
{\frac{1}{2}}{\rho}_{He^{2+}} \notag  \\ 
{\rho}_{b} (1 - f_{H}) & = {\rho}_{He} + {\rho}_{He^{+}} + {\rho}_{He^{2+}}  \notag 
\end{align} \vspace{0.05in}

\noindent
where f$_{H}$ is the primordial hydrogen fraction, m$_{H}$ is the hydrogen mass, and $\rho_{b}$ is the 
baryon density evolved in the original ZEUS-MP hydrodynamics equations.  Any error between the species 
or charge sums and $\rho_{b}$ is assigned to the largest of the species to bring them into agreement with 
$\rho_{b}$.  

Microphysical cooling and heating processes are included by an isochoric operator-split update to the energy 
density computed each time the reaction network is solved: 
\begin{align}
{\dot{e}}_{rad} = k_{ph}\epsilon_{\Gamma}n_{H} & - \Lambda_{rec} n_{H^+} n_{e} - \Lambda_{C} n_{e} \\ 
                & - (\Lambda_{ion} + \Lambda_{exc}) n_{H}n_{e} \notag \label{eqn: egas} 
\end{align} \vspace{0.01in}

\noindent 
where k$_{ph}$ is the photoionization rate described above, $\epsilon_{\Gamma}$ is the fixed energy per 
ionization deposited into the gas (set to 2.4 eV as explained in \citet{wan04}), and $\Lambda_{rec}$, 
$\Lambda_{C}$, $\Lambda_{ion}$, and $\Lambda_{exc}$ are the recombinational, Compton, collisional ionization, 
and collisional excitation cooling rates taken from \citet{anet97}).  These four processes act together with 
hydrodynamics (such as adiabatic expansion or shock heating) to set the temperature of the gas everywhere in 
the simulations.

\subsection{Radiative Transfer}

The radiative transfer module computes k$_{ph}$ by solving the static equation of transfer 
along radial rays outward from a single point source centered in a spherical-polar coordinate 
geometry.  The fact that the medium usually responds to radiation over much longer times than 
the light-crossing times of the problem domain permits us to omit the time derivative in the 
equation of transfer that would otherwise restrict the code to unnecessarily short timesteps.  
This approximation is violated very close to the central star by rapid ionizations that can 
lead to superluminal I-front velocities.  These nonphysical velocities are 
prevented by simply not evaluating fluxes further from the central star than light could have 
traveled by that time in the simulation \citep{anm99}.  Our experience has been that this static 
form of the equation of transfer typically becomes valid before the I-front reaches the Stromgren 
radius, and the code computes very accurate Stromgren radii and formation times.

Photoionizations in any given cell in principle are due to direct photons from the central 
source through the cell's lower face as well as to diffuse recombination photons through all 
its faces.  Recombinations within a zone occur to either the ground state or to any excited
state.  We adopt the on-the-spot (OTS) approximation \citep{o89} that a case (A) recombination 
photon emitted in a zone is reabsorbed before escaping the zone by decomposing both the 
photoionization and recombination terms in the reaction network \vspace{0.1in}  
\begin{align}
\displaystyle\frac{dn_H}{dt} = k^{(A)}n_{H^+} n_{e} & + k^{(B)}n_{H^+} n_{e} - k_{pts}n_{H} \\
                                                    & - k_{diff}n_{H} - k_{coll}n_{H}n_{e} \notag \vspace{0.1in}  
\end{align} \vspace{0.01in}

\noindent
and equating the first and fourth terms, which cancel.  Here k$^{(A)}$ and k$^{(B)}$ are the 
recombination rate coefficients to the ground state and to all excited states while k$_{pts}$ 
and k$_{diff}$ are the rate coefficients for cell ionizations by central stellar photons 
and diffuse photons.  Taking ground-state recombination photons to be reabsorbed before they 
can exit the cell guarantees that no photons enter the cell from other locations in the problem, 
relieving us of the costly radiative transfer from many lines of sight that otherwise would 
be needed to compute the diffuse radiation entering the zone.  While the two terms are set
equal in the equation above, it should be recognized that ionizations typically occur much
more quickly than recombinations.  No error is introduced in the network because the much faster 
photoionizations will simply cycle out any ground-state recombinations over the timestep the 
solution is advanced, regardless of the processes' true timescales.

The OTS approximation is valid anywhere there is a sizeable UV photon mean free path across
a zone, which is the case within the front itself but not usually in the hot ionized gas behind it.
Case (A) photons emitted from an ionized cell might reach the front itself before being absorbed 
because of the very low neutral fraction in their path.  Such photons do not advance the front, 
however, because the neutral atom they leave behind will remove another source or diffuse 
photon that would have reached the front.  Since case (A) recombinations \emph{globally} 
balance diffuse ionizations they can still be thought to cancel in any \emph{single} zone on 
average, even in those that are ionized. 

However, there are important scenarios in which the OTS approximation fails to reproduce the true
ionization of a cosmological structure.  For example, dense gas clumps in primordial minihalos cast 
shadows in the UV field from the central source.  The OTS paradigm would produce shadows that remain 
too sharp by not accounting for the recombination photons that would leak laterally into 
these shadows and soften them.  Failure to capture the correct shadowing may have important 
consequences on the growth of instabilities expected to develop in these ionization fronts 
\citep{gsf96}.  These instabilities are of interest for their potential to promote clumping that 
could later collapse into new stars, especially if mixed with metals from a previous generation.  
Methods for accurate and efficient simulation of recombination radiation are being studied in
connection with 3-D simulations of minihalo photoevaporation under development. 

Only a single integration of the transfer equation along a radial ray from the source at the 
coordinate center is necessary to compute k$_{pts}$ in a cell.  The static transfer equation 
is recast in flux form for simple solution in spherical coordinates:\vspace{0.1in}
\begin{equation}
\nabla \: \cdotp \; {\bf F} \: = \: - \chi\, {\bf F}_r
\end{equation}

\vspace{0.1in}
\noindent 
where $\chi$ is the inverse mean free path of a UV photon in the neutral gas
\begin{equation}
 \chi = \displaystyle\frac{1}{n \sigma}
\end{equation}

\vspace{0.1in}
\noindent 
Simple integration yields the radial flux at the inner face of each zone on the grid: \vspace{0.05in}
\begin{equation}
F_{i} = \left(\displaystyle\frac{r_{i-1}}{r_i}\right)^{2} F_{i-1}\, e^{{-\chi \left(r_{i}-r_{i-1}\right)}} \vspace{0.05in}
\end{equation}
The ionization rate in a zone is calculated in a photon-conserving manner to be 
the number of photons entering the zone per second \vspace{0.15in} 
\begin{align}
n_{ph, inner} & = \displaystyle\frac{F_{i}\, A_{i}}{h\nu}\: \\
              &   \notag \\
              & =\: \displaystyle\frac{F_{i}\,r_{i}^{2}\, \left(
{\phi}_{k+1} - {\phi}_{k}\right)\left(\cos {\theta}_{j} - \cos {\theta}_{j+1} \right)}{h\nu} \notag
\end{align}

\vspace{0.15in}
\noindent minus the number exiting the zone per second \vspace{0.15in}
\begin{eqnarray}
\lefteqn{n_{ph, outer} = \displaystyle\frac{F_{i+1}\, A_{i+1}}{h\nu}\:} \\ \nonumber
 & & \\ \nonumber
 & & = \: \displaystyle\frac{F_{i+1}\,r_{i+1}^{2}\, \left(
{\phi}_{k+1} - {\phi}_{k}\right)\left(\cos {\theta}_{j} - \cos {\theta}_{j+1} \right)}{h\nu}\:  \\ \nonumber
%% & & = \: \displaystyle\frac{\left(\displaystyle
%%\frac{r_{i}}{r_{i+1}}\right)^{2} F_{i}\, e^{{-\chi \left(r_{i+1}-r_{i}\right)}}\,r_{i+1}^{2}\, \left(
%%{\phi}_{k+1} - {\phi}_{k}\right)\left(\cos {\theta}_{j} - \cos {\theta}_{j+1} \right)}{h\nu}\:  \\ \nonumber
%% & & \\ \nonumber
 & &= \: \displaystyle\frac{F_{i}\, e^{{-\chi 
\left(r_{i+1}-r_{i}\right)}}\,r_{i}^{2}\, \left({\phi}_{k+1} - {\phi}_{k}\right)\left(\cos {\theta}_{j} - 
\cos {\theta}_{j+1} \right)}{h\nu}\: \nonumber
\end{eqnarray}

\vspace{0.15in}
\noindent Hence, the ionization rate in the cell is \vspace{0.15in}
\begin{align}
%\lefteqn{n_{ph, inner} - n_{ph, outer}\, =\, n_{ioniz}} \\ \nonumber
% & & \\ \nonumber
% & & =\,  \displaystyle\frac{F_{i}\, \left( 1 - e^{{-\chi 
n_{ioniz} & =\,F_{i}\, \left( 1 - e^{{-\chi \left(r_{i+1}-r_{i}\right)}}\right)\ \\
          &   \notag \\  
          & \times \:  \displaystyle\frac{r_{i}^{2}\, \left({\phi}_{k+1} - {\phi}_{k}\right)\left(\cos {\theta}_{j} - 
\cos {\theta}_{j+1} \right)}{h\nu} \notag
\end{align}

\vspace{0.15in}
\noindent n$_{ioniz}$ must be converted into the rate coeficient required by the chemistry equations.   
In the ionization term of the reaction equation \vspace{0.05in}
\begin{equation}
{\dot{n}}_{H} = -\,k_{ph}\,n_{H} \vspace{0.05in}
\end{equation}
k$_{ph}$ is the ionization rate coefficient, ${\dot{n}}_{H}$ is the number of ionizations per volume
per second, and n$_{ioniz}$ is the number of ionizations per \textit{zone} per second, with the three 
being related by \vspace{0.20in}
\begin{align}
{\dot{n}}_{H} & = \displaystyle\frac{\# \ ionizations}{vol \cdot sec}  \\  
              &   \notag \\
              & = \displaystyle\underbrace{\frac{\# \  atoms}{vol}}_{\displaystyle n_{H}} \ \times\ \underbrace{\displaystyle\frac{\# \ ionizations}{zone \cdot sec} \ \times \ \displaystyle\frac{zone}{\# \  atoms}}_{\displaystyle k_{ph}} \notag
\end{align}
n$_{ioniz}$ is therefore converted to k$_{ph}$ by \vspace{0.15in}
\begin{equation}
k_{ph} = \underbrace{\displaystyle\frac{\# \ ionizations}{zone \cdot sec}}_{\displaystyle n_{ioniz}} \ \times \ 
         \underbrace{\displaystyle\frac{zone}{\# \  atoms}}_{\displaystyle \frac{1}{n_{H}\,V_{cell}}} \label{eqn: kphderiv} \vspace{0.1in}
\end{equation}
\noindent He I and II ionization coefficients as well as Lyman-Werner dissociation rates are also evaluated 
by this procedure.

Our prescription for generating k$_{ph}$ is photon-conserving in that the number of photons emitted along 
any line of sight will always equal the number of photoionizations in that direction over any time interval.   
This formulation enables the code to accurately advance I-fronts with significantly less sensitivity to resolution 
than methods which solve the transfer equation to compute a zone-centered flux or intensity to determine 
k$_{ph}$:\vspace{0.1in}
\begin{equation}
k_{ph} = 4 \pi {\sigma}_{ph} \frac{I(\nu)}{h\nu} d\nu \vspace{0.1in}
\end{equation}
This non-conservative formalism does not guarantee n$_{\gamma}$ to equal n$_{ioniz}$ along a line of 
sight except in the limit of very high resolution.  Such methods can require hundreds or thousands of
radial zones to converge to proper I-front evolution, making photon conservation a very desirable 
property \citep{anm99}. 

However, photon-conserving methods are not necessarily resolution independent, as implied at times 
in the literature.  For example, if a grid fails to resolve a density peak then even a 
photon-conserving algorithm will overestimate the advance of the front (even in a static problem) 
because of the strong n$^{2}$ dependence of recombination rates on material densities.  The resolution
needed for correct I-front evolution in hydrodynamical simulations is governed by the resolution 
necessary for the gas dynamics to converge, since this determines the accuracy of the densities 
encountered by the front.  Nevertheless, photon-conservative schemes are still the method of choice in
gas dynamical I-front simulations because the grid resolution non-conservative methods would require 
to follow the front can be much greater than the resolution needed just for hydrodynamic convergence.

\subsection{Network Solution Schemes}

Each equation in the reaction network can be rewritten as \vspace{0.05in}
\begin{equation}
\frac{dn_{i}}{dt} = C_{i}(T,n_{j}) - D_{i}(T,n_{j}) n_{i}\vspace{0.05in}
\end{equation}
\noindent where C$_{i}$ is the source term representing the formation of species i while D$_{i}$ is 
the source term describing its removal.  A fully-explicit finite-differenced solution would be
 \vspace{0.1in}
\begin{equation}
{n_{i}}^{t + \Delta t} = \frac{C^{t} \Delta t + {n_{i}}^{t}}{1 + D^{t} \Delta t} \vspace{0.1in}
\end{equation}
with C$_{i}$ and D$_{i}$ (and the n$_{j}$ comprising them) being evaluated at the current time.  A fully-
implict update \vspace{0.05in}
\begin{equation}
{n_{i}}^{t + \Delta t} = \frac{C^{t + \Delta t} \Delta t + {n_{i}}^{t}}{1 + D^{t + \Delta 
t} \Delta t} \vspace{0.05in}
\end{equation}
would take the source terms to be at the advanced time, requiring iteration to convergence for the
n$_{i}$

The chemistry rate coefficients driving the reaction network exhibit a variety of different timescales
that make the network numerically stiff.  The shortest reaction times in the network would force a 
purely explicit solution into much smaller timesteps than accuracy requires, while a 
fully-implicit scheme best suited to the solution of stiff networks demands matrix iterations of 
excessive cost in a 3D simulation.  The additional costs of matrix iteration in implicit schemes can 
be offset by the much longer timesteps they permit because the stability of the network is freed from 
its shortest timescale. 

However, a disadvantage of fully-implicit approaches is that they require simultaneous solution of the reaction 
network together with the radiative transfer equation (needed for the ionization rates in the network) 
and the isochoric energy equation (which sets the temperatures utilized in the rate coefficients), all 
evaluated at the advanced time.  While such methods evolve ionization fronts with very high accuracy in 
1D \citep{tet86,ftb90}, photon conservation is sacrificed in the process of achieving concurrency between 
the network, energy, and transfer equations at the future time.  Hence, these algorithms may only achieve their 
superior accuracy if high problem resolutions are employed (8000 radial zones in the \citet{ftb90} theory
validation runs).  Photon-conserving fully-implicit stencils which can operate accurately at much lower 
resolutions may be possible with further investigation.

Furthermore, while in principle a fully-implicit network can be evolved over an entire hydrodynamical 
time, this strategy would accrue significant errors in many density regimes.  \citet{tet86} reported 
that in some test cases it was necessary to evolve their network by no more than a few photoionization 
timescales in order to compute energy deposition correctly.  In such instances the additional costs of 
iteration in implicit schemes outweighs their advantage in accuracy over explicit methods because they 
must ultimately both perform a comparable number of cycles to complete a problem.  It should also be 
noted that highly nonlinear and nonmonotonic primordial heating/cooling rates have been observed to 
retard or prevent Newton-Raphson convergence in implicit cosmological calculations \citep{anet97}. 

We instead adopt the intermediate strategy of sequentially computing each n$_{i}$, building the source 
and sink terms for the i$^{th}$ species' update from the i - 1 (and earlier) updated species while 
applying rate coefficients evaluated at the current problem time \citep{anet97}.  The order of the 
updates is H, H$^{+}$, He, He$^{+}$, He$^{2+}$, e$^{-}$, H$^{-}$, H$^{+}_{2}$, and H$_{2}$.  This 
approach allows direct solution of the densities with sufficient accuracy to follow I-fronts in most 
density regimes with reasonable execution times, which are sometimes much shorter than for implicit 
schemes.  Anninos \etal found a speedup of ten in sequential species updates over an implicit stiff 
solver package in cosmological test cases involving the steady buildup of IGM UV fluxes from a 
metagalactic background.

\begin{figure*}
\epsscale{1.11}
\plottwo{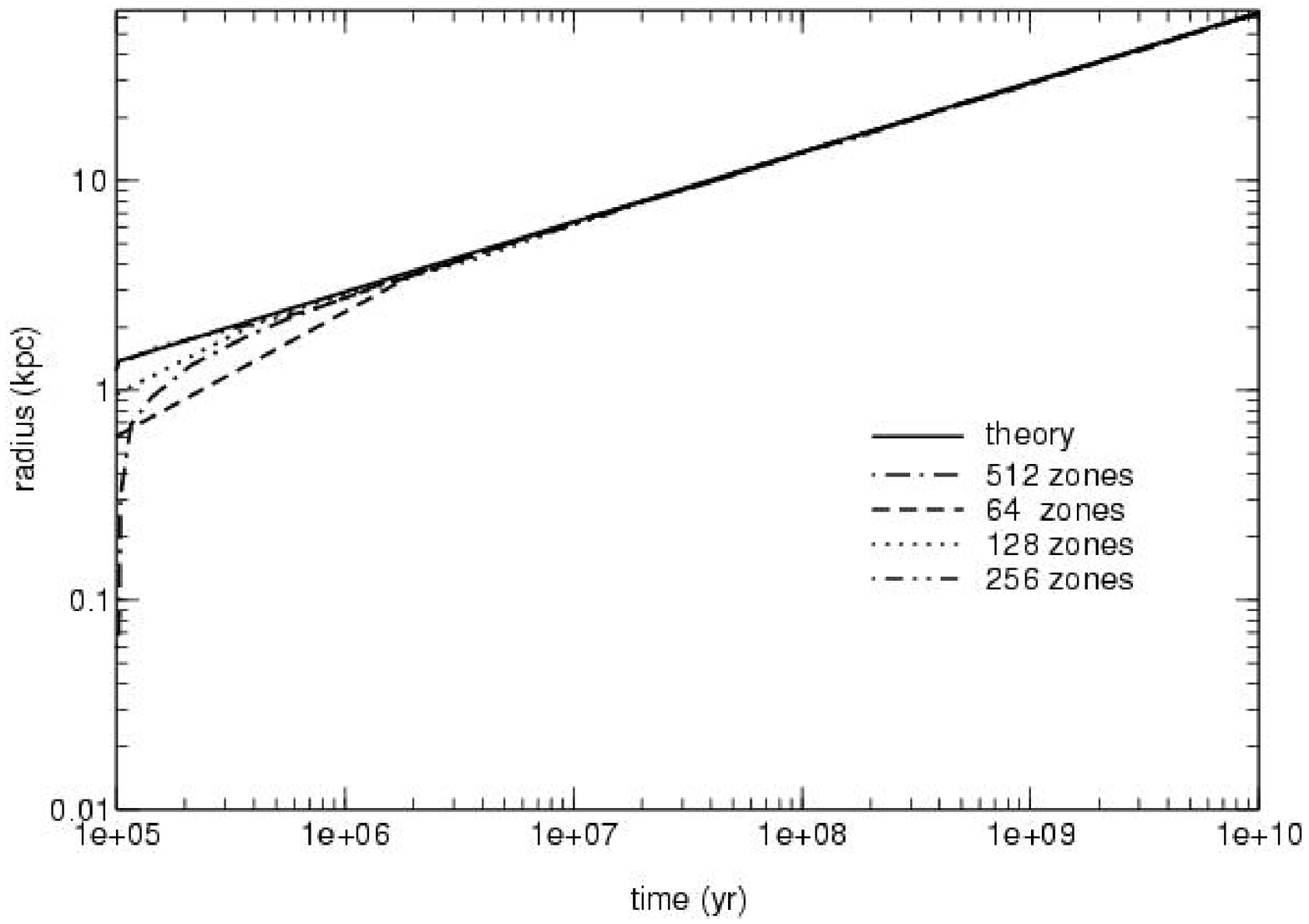}{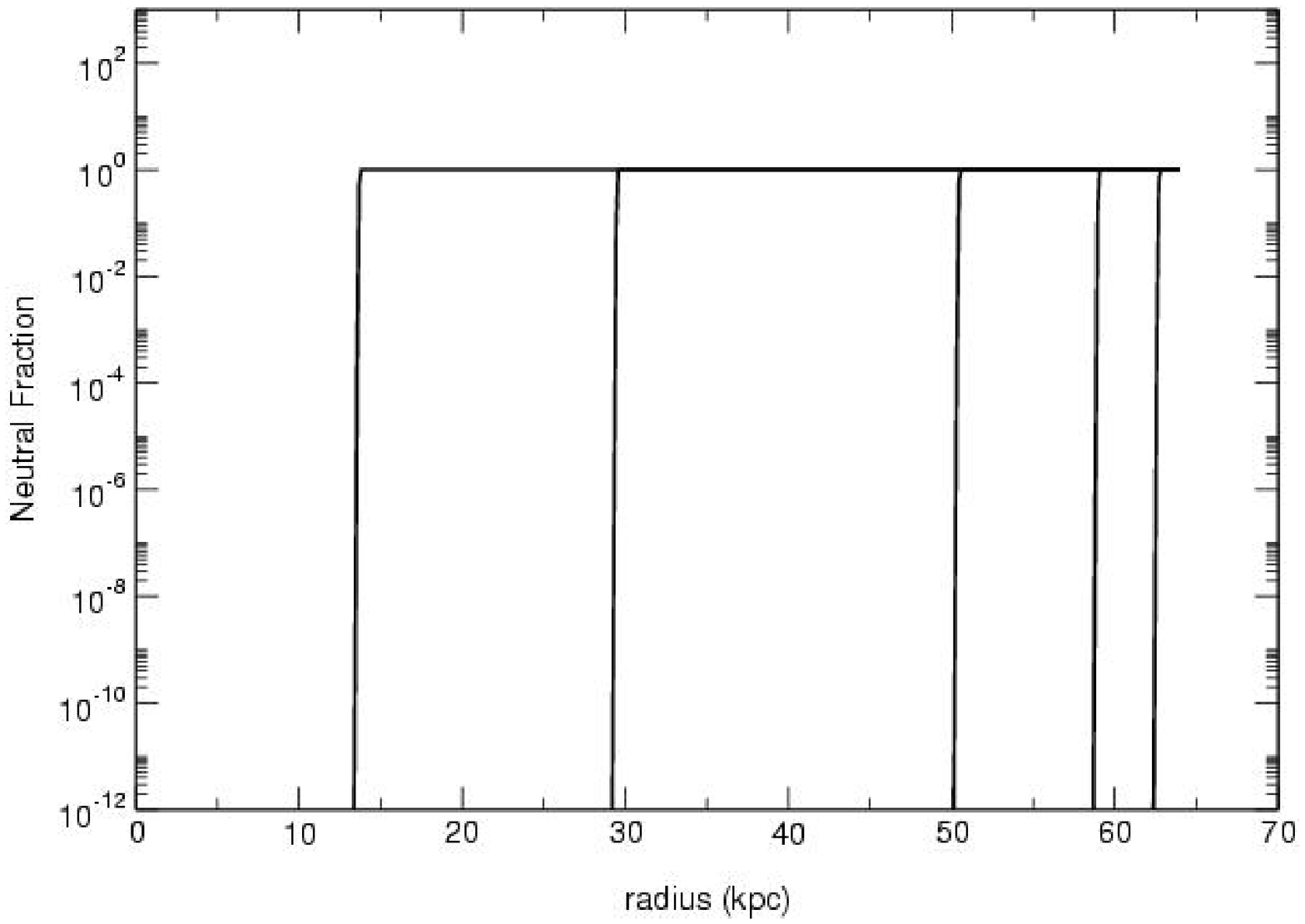}
\caption{R-type ionization front propagating from a point source in a uniform static medium with no 
recombinations.  Left panel:  I-front position as a function of time--theory and algorithm for 64, 
128, 256, and 512 radial zones.  Right panel: neutral fraction profiles--100 Myr, 1 Gyr, 5 Gyr, 8 Gyr, 
and 9.7 Gyr.
\label{fig:T1 radii/ioniz}}  \vspace{0.1in}
\end{figure*}

\subsection{Timestep Control}

Two timescales in general govern the evolution of H II regions.  The many reaction rate timescales can be
consolidated into a single chemistry timestep defined by \vspace{0.1in}
\begin{equation}
t_{chem} = \displaystyle\frac{n_{e}}{{\dot{n}}_{e}}\vspace{0.1in}
\end{equation}
formulated to ensure that the fastest reaction operating at any place or time in the problem determines
the maximum time by which the reaction network may be accurately advanced.  The second timescale is the 
heating/cooling time t$_{heat}$ \vspace{0.1in}
\begin{equation}
t_{heat} = \displaystyle\frac{e_{gas}}{{\dot{e}}_{rad}}\vspace{0.1in} \label{eqn: theat}
\end{equation}
connected to the hydrodynamic response of the gas to the reactions.  The ratio of the two times
can depend on the evolutionary phase of the H II region or even on the current ionization state in a
single zone.  In general, reaction times are shorter than heating 
times as the ionization front propagates outward from the central UV source but cooling times can become 
shorter than recombination times after shutdown of the central source.  The latter circumstance can lead 
to nonequilibrium cooling in H II recombination regions, which can remain much more ionized at low
temperatures than would be expected for a gas in thermodynamic equilibrium, an effect which has been 
observed in cosmological H II region simulations \citep{rgs01,oet05}. 

Many strategies have been devised to interleave reaction networks and radiative transfer
with hydrodynamics.  After computing the global minimum Courant timestep for the
problem domain, Anninos \etal evolve the species in each cell by advancing the rate equations by a tenth 
of the lesser of the 
chemistry and heating/cooling timescales for that cell until a tenth of the cell's heating/cooling timescale 
is covered.  At this point the cumulative energy gained or lost over the chemistry updates is added to 
the cell's gas energy in the microphysical heating/cooling substep described earlier, but neither  
velocities nor densities are updated.  This cycle is then repeated over consecutive heating timesteps 
in the cell until the global minimum Courant timestep is covered.  Cells with the fastest kinetics require 
the most chemistry subcycles over a heating time: more slowly reacting cells covering their heating time 
with fewer subcycles are quiescent during the subsequent cycles required by the faster cells.  Likewise, 
kinetics and energy updates in cells traversing the global hydrodynamical time in fewer heating cycles are halted 
over the additional subcycles the more quickly heating or cooling cells demand.  Every cell in the grid 
undergoes the same number of subcycles (which continue until the last cell has covered the global Courant
timestep), but updates in a given cell are suspended after it has crossed this timestep.  New photoionization
rates are calculated every chemistry timestep by a call to a radiative transfer module but the other rate 
coefficients remain constant over a heating timestep because they depend only on temperature; they are updated 
at the beginning of the next heating cycle with the new gas temperature.  At the end of the hydrodynamical 
timestep full source and advective updates of velocities, energies, and conserved total baryonic densities
are performed.

Although sufficient for the slowly rising UV metagalactic backgrounds in the Anninos \etal calculations this
subcycling approach does not accurately simulate the growth of ionization fronts.  Proper
front capturing is sensitive to velocities that can build up over the heating timestep that are not correctly
computed by the Anninos \etal scheme because it does not update velocities until many such heating times have passed.  
The order of execution of our algorithm is as follows: first, the radiative transfer module 
is called to compute k$_{ph}$ via eq. (\ref{eqn: kphderiv}) to determine the smallest heating/cooling timestep on the grid.  The grid minimum
of the Courant time is then calculated and the smaller of the two is adopted as t$_{hydro}$; in a 1--D calculation
this is 
\begin{align}
t_{hydro} & = min(t_{cour}, t_{heat}) \\
          &        \notag \\
          & = min \left(\displaystyle\frac{\Delta\,r}{c_{s}},\, 0.1 \, 
            \displaystyle\frac{e_{gas}}{{\dot{e}}_{rad}} \right) \notag
\end{align}
Next, from this same set of k$_{ph}$ the shortest chemistry timestep of the grid is calculated. \vspace{0.05in}
\begin{equation}
t_{chem} = 0.1 \, \displaystyle\frac{n_{e}}{{\dot{n}}_{e}} \vspace{0.05in}
\end{equation}
The species densities and gas energy in \emph{all} cells are then advanced over this 
timestep, the transfer module is called again to compute a new chemistry timestep, and the 
network and energy updates are performed again.  The n$_{i}$ and energy are subcycled over 
successive chemistry timesteps until t$_{hydro}$ has been covered, at which point full source 
and advective updates of velocities, energies, and total densities are performed.  A new t$_{hydro}$ 
is then determined and the cycle repeats.

\begin{figure*}
\epsscale{1.11}
\plottwo{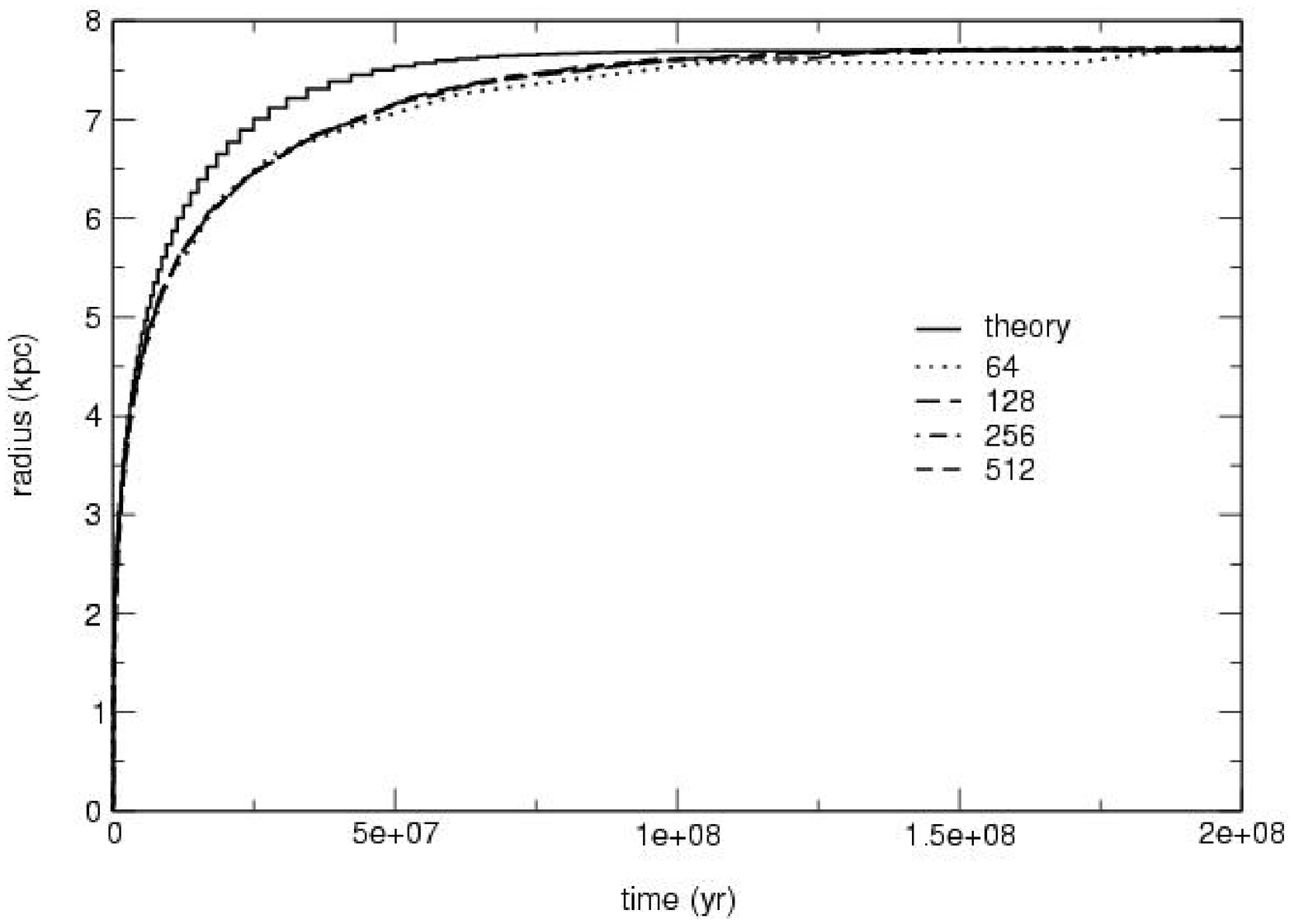}{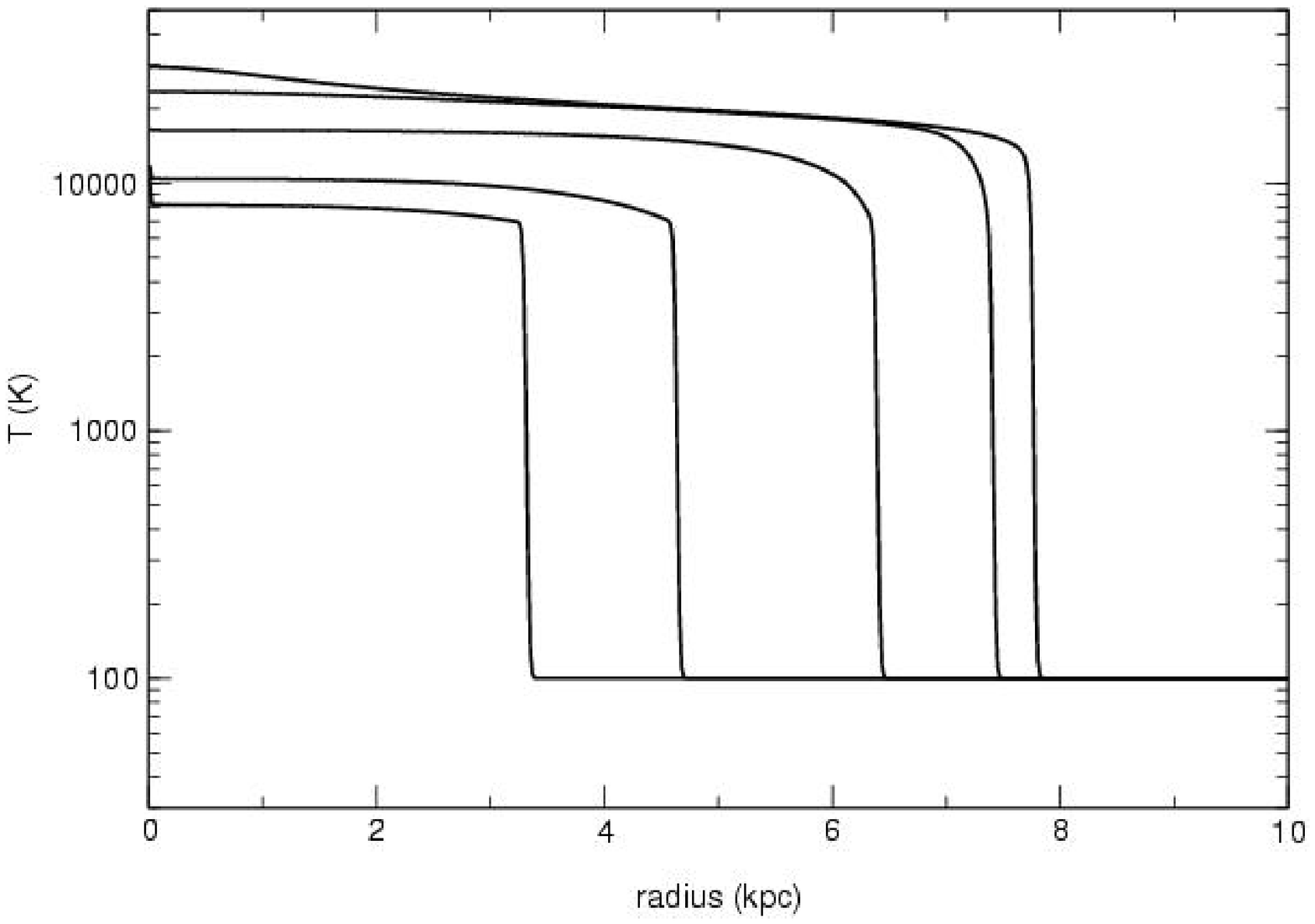}
\caption{R-type ionization front propagating from a point source in a uniform static medium with recombinations.  Left panel:
I-front position as a function of time--theory and algorithm for 64, 128, 256, and 512 radial zones.  Right panel: temperature
profiles for a 512 radial zone simulation at 1.77 Myr, 5.31 Myr, 21.2 Myr, 63.7 Myr, and 173.4 Myr.
\label{fig:T2 radii/temp}} \vspace{0.1in}
\end{figure*}

If the chemistry timestep is longer than the global hydrodynamical time the reaction network
is only subcycled once.  This more restrictive hierarchy of rate solves and hydrodynamical
updates is necessary to compute the correct velocities in each zone with the passage of the 
front over the wide range of density regimes discussed in section 3.  Note that cooling times 
which are shorter than recombination times in the problem are easily handled because 
the code will simply cycle the reactions in each cell once over the hydrodynamic timestep.
Considerable experimentation with alternative hierarchies of kinetic and hydrodynamical updates 
and choices of timestep control (some involving photoionization times) proved them to be less 
accurate or robust. 

\section{Ionization Front Physics in Power-Law Densities}

\citet{ftb90} performed analytical studies of 1-D ionization fronts from a monochromatic source of photons centered in a radial 
density profile with a flat central core followed by an r$^{-\omega}$ 
dropoff:\vspace{0.1in}
\begin{equation}
n_{H}(r) = 
\begin{cases}
n_{c} & \text{if r $\leq$ r$_{c}$} \\
n_{c}(r/r_{c})^{-\omega} & \text{if r $\geq$ r$_{c}$}
\end{cases} \vspace{0.1in}
\end{equation}
They considered photon rates that would guarantee that the Stromgren radius R$_{S}$ of the front if the 
entire medium was of uniform density n$_{c}$ is greater than r$_{c}$, noting that if R$_{S}$ $<$ r$_{c}$ 
that the front would evolve as if it were in a constant density.  Their analysis of I-front propagation down 
the radial density gradients revealed that there is a critical exponent
\vspace{0.1in}
\begin{equation}
\omega_{crit}=\frac{3}{2}\left[1-\left(\frac{r_{c}}{R_{S}}\right)^{3}\right]^{-1} \label{eqn: wcrit} \vspace{0.1in} 
\end{equation}
below which the front executes the classic approach to a Stromgren sphere of modified radius \vspace{0.20in} 
\begin{equation}
R_{\omega} = R_{S}\left[\frac{3-2\omega}{3} + \frac{2\omega}{3}\left({\frac{r_{c}}{R_{S}}}
\right)^{3}\right]^{1/(3-2\omega)}\left(\frac{R_{S}}{r_{c}}\right)^{2\omega/(3-2\omega)}\vspace{0.20in}
\end{equation}
at which point it reverts from R-type to D-type and continues along the gradient, building up a dense shocked
shell before it.  Here R$_{S}$ is the classical Stromgren radius the ionization front would have in a uniform
density medium \vspace{0.1in}
\begin{equation}
R_S = \left(\displaystyle\frac{3 {\dot{n}}_{ph}}{4 \pi {\alpha}_B {n_{c}^2}} \right)^{1/3}\vspace{0.1in}
\end{equation}
If $\omega$ $<$ 1.5 $<$ ${\omega}_{crit}$ the front remains D-type throughout its lifetime and 
continues to accumulate mass in its shell, expanding as \vspace{0.1in}
\begin{equation}
R(t) = R_{\omega}\left[1 + \frac{7-2\omega}{4} + \left(\frac{12}{9-4\omega}\right)^{1/2}
\frac{c_{i}t}{R_{\omega}}\right]^{4/(7-2\omega)} \label{eqn: r_w1} \vspace{0.10in}
\end{equation}
where c$_{i}$ is the sound speed in the ionized gas.  If $\omega$ = 1.5 the shock and front coincide without
any formation of a thin neutral shell as in the $\omega$ $<$ 1.5 profiles.  When 1.5 $<$ $\omega$ $<$ ${\omega}_{crit}$ 
the D-type front will revert back to R-type and quickly overrun the entire cloud.  Since I-fronts ultimately 
transform back to R-type in any cloud with density dropoffs steeper than 1.5 this power constitutes the critical 
point for eventual runaway ionization.  As expected, if $\omega$ = 0 then R$_{\omega}$ becomes R$_{S}$ and R(t)
exhibits the t$^{4/7}$ expansion of an ionization front in a uniform medium \citep{o89}.

Fronts descending gradients steeper than r$^{{-\omega}_{crit}}$ never slow to a Stromgren radius or transform
to D-type.  Remaining R-type, they quickly ionize the entire cloud, leaving behind an essentially undisturbed 
ionized density profile because they exit on timescales that are short in comparison to any hydrodynamical 
response of the gas.  Completely ionized and at much higher pressures, the entire cloud begins to inflate outward 
at the sound speed of the ionized material.  However, the abrupt core density dropoff left undisturbed by the rapid 
exit of the front develops a large pressure gradient because of the equation of state in the nearly isothermal 
postfront gas.  The sharpest pressure gradient is at the ionized core's edge at r $\sim$ r$_{c}$.  This edge expands 
outward in a pressure wave which quickly steepens into a shock that overtakes the more slowly moving outer cloud 
regions.  The velocity of this shock depends on the initial density
dropoff: if $\omega_{crit} < \omega <$ 3 then \vspace{0.10in}
\begin{equation}
r_{c}(t) \simeq r_{c} + \left[1+\left({\frac{3}{3-\omega}}\right)^{1/2}\right]c_{i}t \label{eqn: om2} \vspace{0.1in}
\end{equation}
If $\omega$ = 3 then 
\begin{equation}
r_{c}(t) \simeq 3.2\,r_{c}\left({\frac{c_{i}t}{r_{c}}}\right)^{1.1} \label{eqn: om3}
\end{equation}
and if $\omega > $ 3 \vspace{0.10in}
\begin{equation}
r_{c}(t) \simeq r_{c}\left[1+\left({\frac{4}{\omega-3}}\right)^{1/2}\left({\frac{\delta+2-\omega}{2}}\right)\left({\frac{c_{i}t}{r_{c}}}\right)\right]^{2/\left(\delta+2-\omega\right)} \label{eqn: om5} \vspace{0.1in}  
\end{equation}
\noindent where $r_{c}$ is the initial core radius and $\delta$ is an empirically fit function of $\omega$: \vspace{0.1in}
\begin{equation}
\delta \simeq 0.55\, \left(\omega-3\right) \: + \: 2.8 \vspace{0.1in}
\end{equation}
\noindent The shock has a constant velocity for $\omega_{crit}$ $<$ $\omega$ $\leq$ 3, weakly accelerates for $\omega$ = 3, and  
strongly accelerates if $ \: \omega >  3$.  This is in agreement with what would be expected for the mass of each cloud.  For
$\omega$ $<$ 3 the cloud mass is infinite so a central energy source could not produce gas velocities that increase with time.
The cloud mass becomes finite just above $\omega$ = 3, the threshold for the ionized flow to exhibit a positive acceleration.

\section{Algorithm Tests: Static Media}

% 
%%\begin{figure*}
%%\epsscale{0.65}
%%\plotone{f3.eps}
%%\caption{Ionization profiles for a 512 radial zone simulation at 1.77 Myr, 5.31 Myr, 21.2 Myr, 63.7 Myr, and 173.4 Myr. 
%%\label{fig:T2 ioniz}}
%%\end{figure*}
% 
 
We present a series of static I-front tests of increasing complexity for initial code validation.  In line with the 
non-hydrodynamical nature of these problems, velocity updates were suspended but the heating and cooling updates in eq 
(\ref{eqn: egas}) were performed.  The energy updates were necessary to evolve the reaction rates according to temperature 
as well as to regulate the timesteps over which the reaction network was advanced.  The initial gas temperature in all the
static tests was set to 10 K.

\subsection{Spherical I-fronts in Uniform Media}

\begin{figure}
\resizebox{3.45in}{!}{\includegraphics{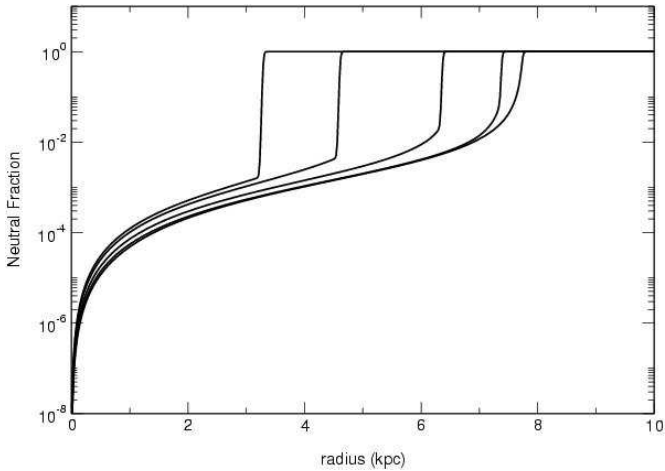}}
\caption{Ionization profiles for a 512 radial zone simulation at 1.77 Myr, 5.31 Myr, 21.2 Myr, 63.7 Myr, and 173.4 Myr.} \vspace{0.25in}
\label{fig:T2 ioniz}
\end{figure}

The simplest test is an R-type ionization front due to a monochromatic point source in a uniform infinite static medium in which no 
recombinations occur \citep{anm99}.  The radius of the spherical front is easily computed by balancing the number of emitted photons 
with the number of atoms in the ionized volume: \vspace{0.05in}
\begin{equation}
R(t) = \left(\displaystyle\frac{3 {\dot{n}}_{\gamma}}{4 \pi {n_{H}}} \; t \right)^{1/3} \vspace{0.05in}
\end{equation}
The front will expand forever in the absence of recombinations but will eventually slow to zero velocity as t $\rightarrow \infty$.  We compare this solution to our code results in Fig \ref{fig:T1 radii/ioniz} for n$_{H}$ 
= 10$^{-2}$ cm$^{-3}$, ${\dot{n}}_{\gamma}$ = 10$^{51}$ s$^{-1}$, and outer boundary of 64 kpc for grid resolutions of 64, 128, 256, 
and 512 radial zones.  The position of the front is defined to be at the outermost zone whose neutral fraction has decreased to 50\%. 
The time evolution of the front is shown in Fig \ref{fig:T1 radii/ioniz}.  For a given resolution the algorithm is within one zone of the exact 
solution after 2.0 $\times$ 10$^{6}$ yr.  As expected for a static homogeneous medium, the photon-conserving radiative transfer correctly propagates the R-type front 
independently of numerical resolution.  The neutral fraction profiles are extremely sharp because there are no recombinations, and they
drop essentially to zero behind the front. 

\begin{figure*}
\epsscale{1.05}
\plottwo{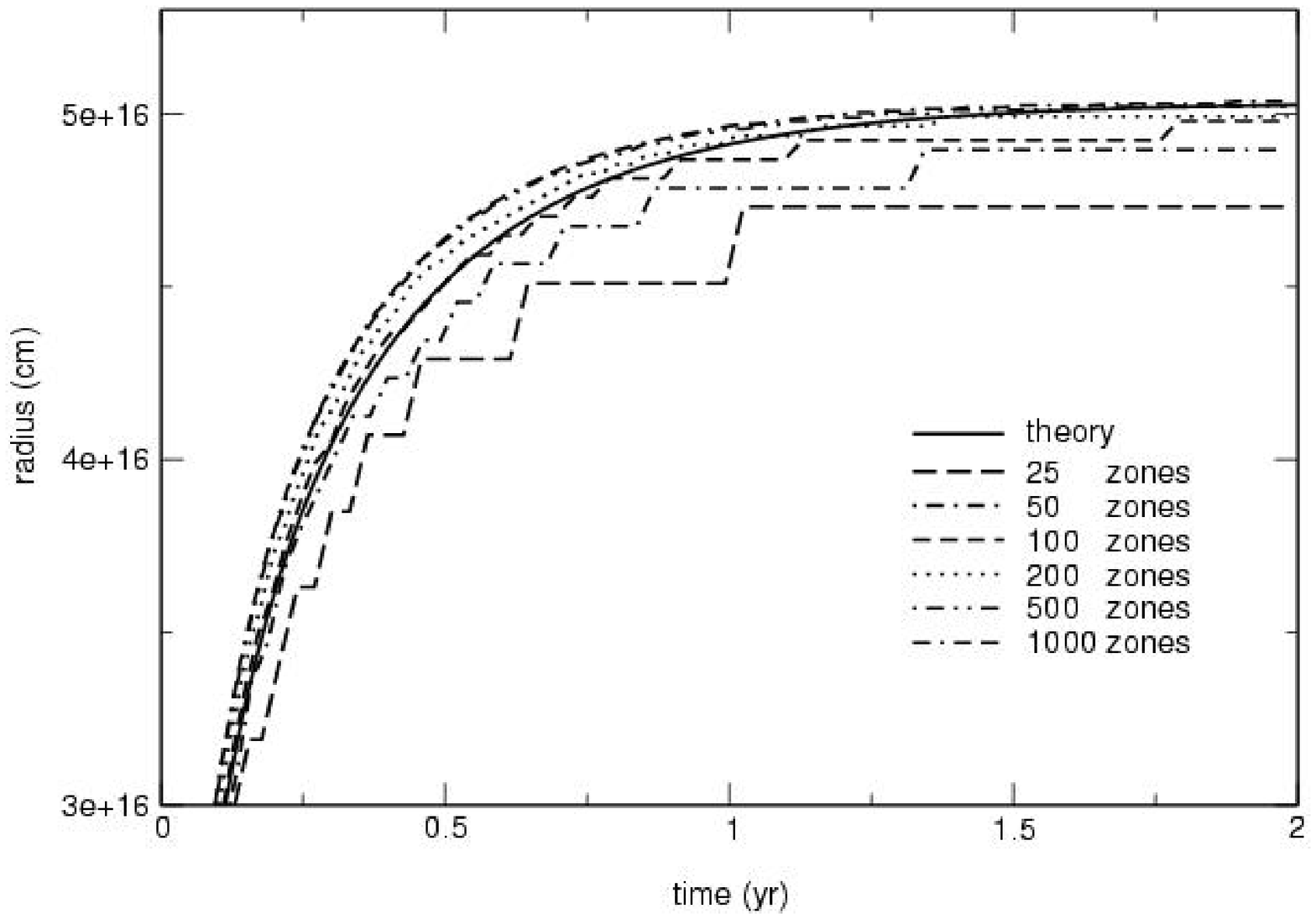}{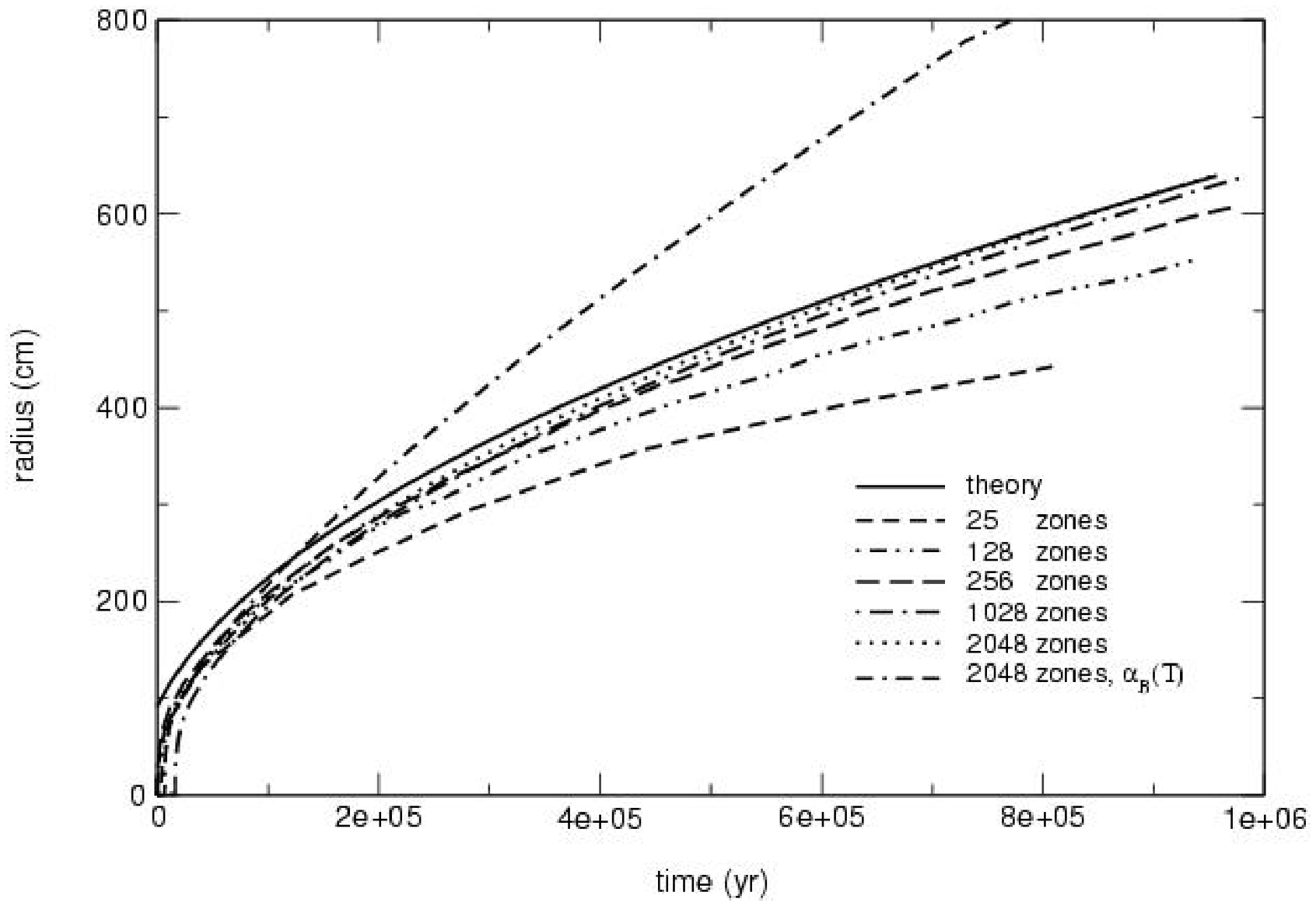}
\caption{Left panel: approach of the I-front to its final Stromgren radius in a static r$^{-1}$ density profile with n$_c$ = 10$^6$ 
cm$^{-3}$, r$_{c}$ = 2.1e16 cm, constant $\alpha_{B}$ = 2.4773e-13, and an outer boundary of 5.5e16 cm for 25, 50, 100, 200, 500, and 
1000 radial zones.  Right panel: I-front position as a function of time in the r$^{-2}$ density profile with n$_c$ = 3.2 cm$^{-3}$, 
r$_{c}$ = 91.5 pc, constant $\alpha_{B}$, and an outer boundary of 0.8 kpc for 64, 128, 256, 1024, and 2048 radial zones.  The uppermost 
curve is for a 2048-zone temperature-dependent $\alpha_{B}$ simulation. \label{fig: w1/w2}}  \vspace{0.1in}
\end{figure*}

Including recombinations yields the well-known result for an I-front in a uniform infinite static medium: \vspace{0.1in}
\begin{equation}
R(t) = R_S \left[ 1 - exp(- t/t_{rec})\right]^{1/3} \label{eqn: r2(t)}\vspace{0.1in}
\end{equation}
where R$_{S}$ is the Stromgren radius \vspace{0.1in}
\begin{equation}
R_S = \left(\displaystyle\frac{3 {\dot{n}}_{ph}}{4 \pi {\alpha}_B(T) {n_{H}^2}} \right)^{1/3} \label{eqn: r_S} \vspace{0.1in}
\end{equation}
and \vspace{0.1in}
\begin{equation}
t_{rec} = [{\alpha}_B(T) n_{H}]^{-1} \vspace{0.1in}
\end{equation}
Taking again n$_{H}$ = 10$^{-2}$ cm$^{-3}$ and ${\dot{n}}_{\gamma}$ = 10$^{51}$ s$^{-1}$ but with an outer radius of 10 kpc, we plot the 
results of our algorithm for 64, 128, 256, and 512 radial zones along with the analytical solution in Fig \ref{fig:T2 radii/temp}.  The computed curves
exhibit excellent agreement with theory, with a maximum error of 7.5\% between the 64-zone solution and eq. (\ref{eqn: r2(t)}).  The code results are
again clustered closely together because of photon conservation, and after several recombination times they converge to a Stromgren 
radius of 7.72 kpc, within 0.3\% of the R$_S$ = 7.70 kpc predicted by eq. (\ref{eqn: r_S}).

The small departure from theory at intermediate times evident in Fig \ref{fig:T2 radii/temp} arises because eq. (\ref{eqn: r2(t)}) assumes a constant $\alpha_B$(T) 
throughout the evolution of the front.  In reality, the H II region has a temperature structure that changes over time, as shown
in Fig \ref{fig:T2 radii/temp}.  At any given time the temperature decreases from its maximum near the point source to its minimum at the I-front, and this 
drop in temperature with radius can grow to more than 10000 K at later times.  We adopt an average temperature of 20000 K for $\alpha_B$ in eq 
 (\ref{eqn: r2(t)}).  The postfront temperatures are greatest near the central star because the gas there has undergone more cycles of recombination 
and photoionization than the gas near the front.  Each cycle increments the gas temperature upward because lower energy electrons are 
preferentially recombined with a net deposition of energy into the gas.  Successive profiles steadily rise in temperature over time for 
the same reason.

The temperature profiles continue to rise well after the Stromgren radius is reached as seen in Fig \ref{fig:T2 radii/temp} because there are no gas 
motions in which PdV work can be performed and because collisional excitation and ionization processes are suppressed by the decline 
of postfront neutral fractions with time.  Neutral fractions fall as rising temperatures slow down recombinations.  The rising gas 
temperatures in this static problem eventually stall when they sufficiently quench recombinations, well before other processes such as
bremmstrahlung cooling arise.  Inclusion of recombinations leads to the ionization structure of this static H II region in Fig \ref{fig:T2 ioniz} (which in part 
is determined by the temperature profile), in contrast with the very simple neutral fraction profiles of the previous test.
\citet{sg87} extend this classical H II region problem to a ionizing point source in a uniform medium undergoing 
cosmological expansion in a Friedmann-Robertson-Walker universe, but this test cannot be performed by ZEUS-MP at
present because scale factors are not implemented in the code. 

\subsection{I-fronts in r$^{-\omega}$ Density Gradients} \label{sect: rmw static}

Franco \etal studied the hydrodynamics of I-fronts in power-law gradients but not their time-dependent propagation in static profiles.  Solutions for R-type fronts exiting flat central cores into r$^{-\omega}$ gradients do exist for static 
media but in general are quite complicated. 

\subsubsection{Bounded I-fronts}

\citet{met05} found that in $\omega$ = 1 Franco \etal static density profiles the radius of the front advances according to 
\vspace{0.1in} 
\begin{equation}
R(t)=R_S\left\{1+{\rm W_0}\left[-\exp\left(-\frac{r_ct}{R_St_{\rm rec,core}}-1\right)\right]\right\} \label{eqn: r4(t)} \vspace{0.1in}
\end{equation}
where R$_S$=L/K is the Stromgren radius and W$_0$(x) is the principal branch of the Lambert W function \citep{cor96}.  W(x) 
is a solution of the algebraic equation $W(x)e^{W(x)}=x$ and must be evaluated numerically.  Here, $L\equiv{\dot{N}_\gamma}/{
(4\pi n_cr_c)}$ and $K\equiv{n_cr_cC\alpha_B}=r_c/t_{\rm rec,core}$, where $t_{\rm rec,core}\equiv(n_cC\alpha_B)^{-1}$ is the 
recombination time in the core and C is the clumping factor.  Eq. (\ref{eqn: r4(t)}) describes the approach of the bounded front to its modified Stromgren 
radius R$_S$ with time.

W(x) in general is multivalued in the complex plane; its principal branch W$_0$(x) is single-valued over the range [-1/e,0], 
monotonically increasing from -1 to 0 over this interval.  As the time in eq. (\ref{eqn: r4(t)}) evolves from zero to infinity the
argument of W$_0$(x) advances from -1/e to 0, guaranteeing that the I-front expands from zero to R$_S$ in approximately twenty
recombination times.  Several commercial algebraic software packages can compute W(x); we instead utilized a recursive algorithm
by Halley \citep{cor96}:\vspace{0.1in}
\begin{equation}
w_{j+1}\: =\: w_j\: - \:\displaystyle\frac{w_j\,e^{w_j} - x}{e^{w_j}\,(w_j + 1) - \displaystyle\frac{(w_j + 2)(w_j\,e^{w_j} - x)}{2(w_j + 1)}}
\vspace{0.1in}
\end{equation}
Adopting the first two terms in the series expansion for W(x) as an initial guess for each value of x \vspace{0.1in}
\begin{equation}
W(x)\: = \: -1\:+\:p\:-\:\frac{1}{3}\,p^{2}\:+\:\frac{11}{72}\,p^{3}\:+\:\cdots \hspace{0.25in} p\: = \: +2\sqrt{2\,(ex+1)}
\vspace{0.1in}
\end{equation}
this method typically converged to an error of less than 10$^{-6}$ in 2 - 4 iterations per point.

\begin{figure*}
\epsscale{1.08}
\plottwo{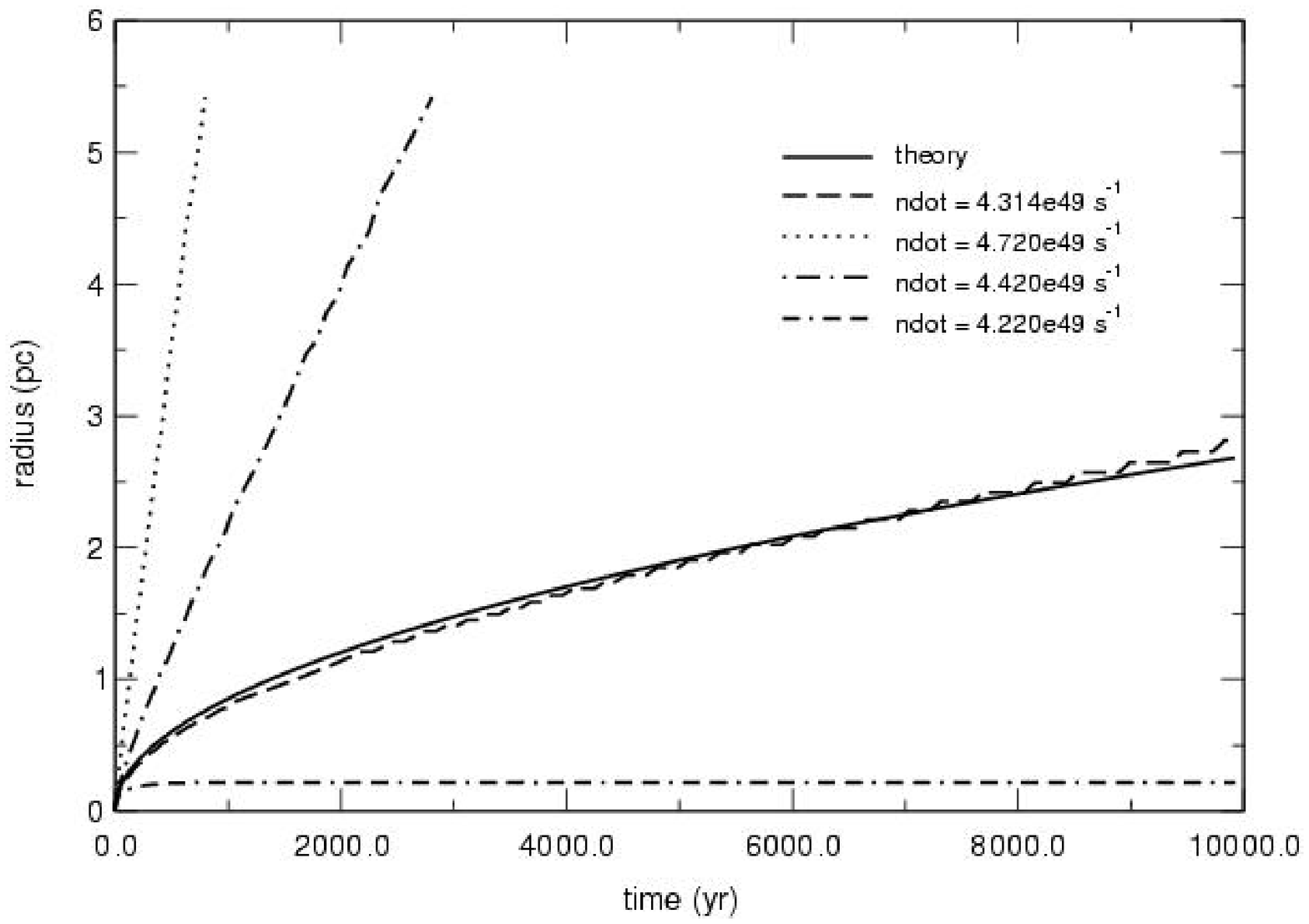}{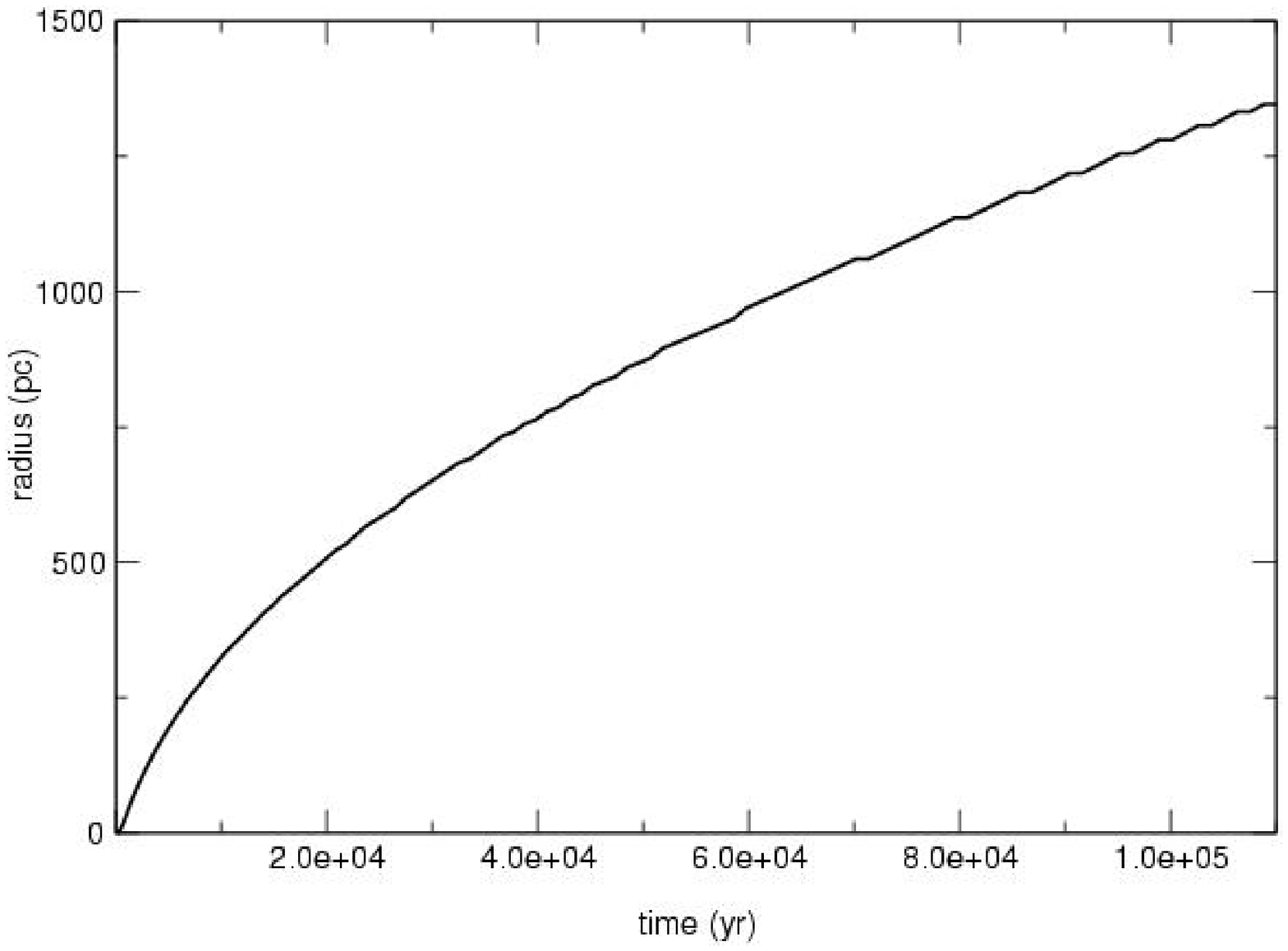}
\caption{Left panel: evolution of the R-type front in the r$^{-2}$ density profile with n$_c$ = 10$^6$ cm$^{-3}$, r$_{c}$ = 2.1 $\times$ 
10$^{12}$ cm, and outer boundary of 5.5 pc for for the central ionizing photon rates shown.  Right panel: R-type I-front position in the same r$^{-2}$ density field extended to an outer boundary of 1500 pc with 500 ratioed zones. \label{fig: w2/w2long}} \vspace{0.1in}
\end{figure*}

We show in the left panel of Fig \ref{fig: w1/w2} the radius of the ionization front on uniform grids of 25,
50, 100, 200, 500, and 1000 zones with inner and outer boundaries of 1.047 $\times$ 10$^{11}$ cm and 5.6 
$\times$ 10$^{16}$ cm, respectively.  We set n$_{c}$ = 10$^{6}$ cm$^{-3}$, r$_{c}$ = 2.1 $\times$ 10$^{16}$ cm, 
$\alpha_B$ = 2.4773 $\times$ 10$^{-13}$ (constant with temperature), and ${\dot{n}}_{\gamma}$ = 5.0 $\times$ 
10$^{49}$ s$^{-1}$, which results in a core recombination time of 0.128 yr and a final Stromgren radius of
5.03 $\times$ 10$^{16}$ cm.  ZEUS-MP converges to within 4\% of eq. (\ref{eqn: r4(t)}) by 500 radial zones at early 
times and to within 1\% past 1.5 yr (on the scale of the graph, the 500 zone curve is essentially identical to
the 1000 zone plot).  The disagreement between theory and simulation clearly illustrates that photon-conserving schemes can fail to properly 
evolve ionization fronts if important density features are not resolved, in this case the abrupt falloff in 
density at the edge of the central core.  Nevertheless, the code still exhibits good agreement ($>$ 10\%) with the 
analytical result with relatively few (50) zones.
\begin{figure*}
\epsscale{1.05}
\plotone{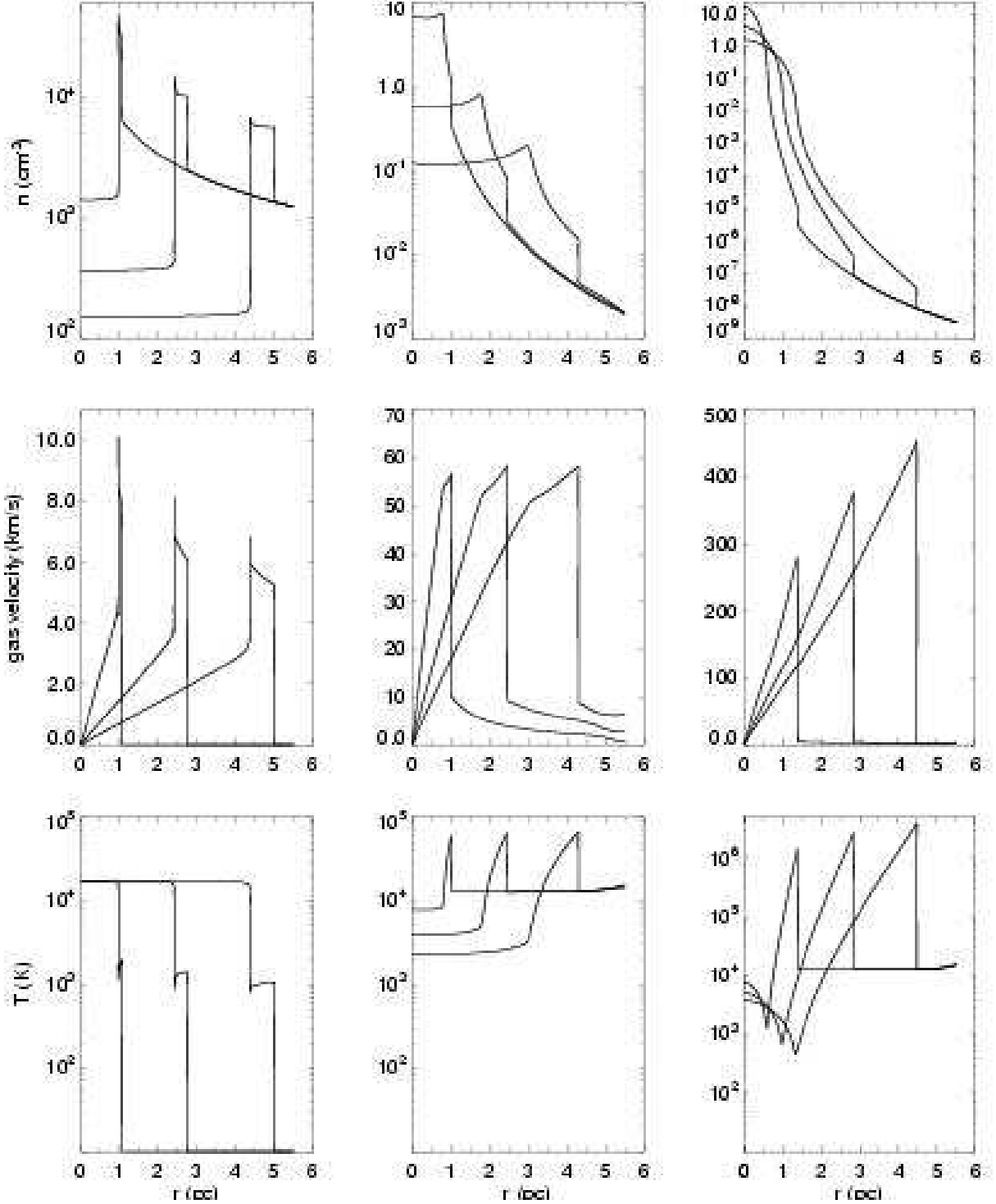}\vspace{0.15in}
\caption{Flow profiles of I-fronts and ionized shocks.  Column 1: density, velocity, and temperature profiles of an 
ionization front advancing along an $\omega$ = 1.0 density gradient. Output times are 8.64e04 yr, 2.73e05 yr, and 
5.71e05 yr.  Column 2: same profiles, but for an ionized core shock in an $\omega$ = 3.0 density gradient. Output 
times are 1.43e04 yr, 3.23e04 yr, and 5.50e04 yr. Column 3: flow profiles of an ionized core shock expanding in an 
$\omega$ = 5.0 density gradient. Output times are 6123 yr, 9380 yr, and 1.23e04 yr. \label{fig: 123profiles}} 
\vspace{0.075in}
\end{figure*}

\subsubsection{Unbounded I-fronts}

The evolution of an R-type front in a static $\omega$ = 2 Franco density field in general involves complex 
Lambert W functions with several branches but reduces to the relatively simple unbounded solution \citep{met05} \vspace{0.1in} 
\begin{equation}
R(t) = r_c \left[ 1 + 2t/t_{rec,core}\right]^{1/2} \label{eqn: r3(t)}\vspace{0.1in}
\end{equation}
provided that \vspace{0.1in}
\begin{equation}
{\dot{n}}_{\gamma} = \frac{16\pi}{3}\: {r_c}^{3} {n_c}^{2} {\alpha}_B \label{eqn: ndot} \vspace{0.05in}
\end{equation}
We plot the position of the ionization front in the right panel of Fig \ref{fig: w1/w2} on a uniform grid for the resolutions indicated, with outer radius 
0.8 kpc, n$_{c}$ = 3.2 cm$^{-3}$, r$_{c}$ = 91.5 pc, and a constant $\alpha_B$ = 2.4773 $\times$ 10$^{-13}$, which yields 
a recombination time in the core of 0.04 Myr and ${\dot{n}}_{\gamma}$ = 9.55 $\times$ 10$^{50}$ s$^{-1}$.  The \citet{met05} 
solutions assume a specific temperature for the H II region that does not evolve with time, so $\alpha_B$ was set constant 
in our simulation.  Our ZEUS-MP results converge to within 10\% of eq. (\ref{eqn: r3(t)}) by 2000 radial zones (and are within 2 
- 3\% over most of 
the time range) with ${\dot{n}}_{\gamma}$ = 9.65 $\times$ 10$^{50}$ s$^{-1}$.  An interesting aspect of the analytical solution is 
its sensitivity to the requirement that $\alpha_B$ is constant, as seen in the $\alpha_B$(T) plot in the right panel of Fig 
\ref{fig: w1/w2}.  Recombinations are supressed in a static H II region whose temperature varies with radius and increases with time, 
freeing the front to advance more quickly than expected from eq. (\ref{eqn: r3(t)}).

R(t) is also strongly divergent from eq. (\ref{eqn: r3(t)}) for photon rates above or below eq. (\ref{eqn: ndot}), as seen in the left panel of Fig \ref{fig: w2/w2long}.  We set n$_{c}$ = 
10$^{6}$ cm$^{-3}$, r$_{c}$ = 2.1 $\times$ 10$^{16}$ cm, and $\alpha_B$ = 2.4773 $\times$ 10$^{-13}$ in a simulation with 
a 172-zone ratioed grid with an outer radius of 5.5 pc.  These much higher and more compact central densities were chosen 
for their relevance to high-redshift cosmological minihalos and are more computationally demanding.  The ratioed grid is 
defined by the requirements \vspace{0.1in}
\begin{equation}
\displaystyle\frac{\Delta r_{i+1}}{\Delta r_{i}} = \beta \hspace{0.25in} r_{n} - r_{1} = r_{outer} - r_{inner} \vspace{0.1in}
\end{equation}
where n is the number of radial zones and r$_{inner}$ was chosen to be small in comparison to 5.5 pc but large enough to
avoid coordinate singularities at the origin.  We applied a grid ratio $\beta$ = 1.03 to concentrate zones at the origin 
in order to resolve the central core and density drop.  Our choice of parameters sets ${\dot{n}}_{\gamma}$ = 3.844 $\times$ 
10$^{49}$ s$^{-1}$ and t$_{rec,core}$ = 0.128 yr.  As shown in Fig \ref{fig: w2/w2long}, the ionization front in our calculation stalls for 
this value of ${\dot{n}}_{\gamma}$ but agrees with eq. (\ref{eqn: r3(t)}) to within 5\% if we change ${\dot{n}}_{\gamma}$ to 4.314 $\times$ 
10$^{49}$ s$^{-1}$, a difference of 12\%.  If the photon rate is increased another 2\% the front exits the grid much faster 
than predicted by eq. (\ref{eqn: r3(t)}) and if it is decreased by 2\% the front is halted, implying that the rate set by eq. (\ref{eqn: ndot}) 
is the threshold for breakout from an r$^{-2}$ core envelope. 

This result is confirmed by substituting eq. (\ref{eqn: ndot}) into eq. (\ref{eqn: r_S}) to compute the R$_S$ appearing in eq. (\ref{eqn: wcrit}), 
which yields $\omega_{crit}$ = 2.  This gradient is just steep enough for the central flux of eq. (\ref{eqn: ndot}) to be unbounded, with the 
position of the front evolving according to the power-law eq. (\ref{eqn: r3(t)}).  We note that while the front escapes for $\omega$ $>$ 
$\omega_{crit}$, it will slow down for $\omega$ $<$ 3 because the cloud mass is infinite but approach the speed of light for $\omega$ $>$ 3 
because of finite cloud mass.  The eventual slowing of the front in an $\omega$ = 2 cloud for ${\dot{n}}_{\gamma}$ = 5.0 $\times$ 
10$^{49}$ $>$ ${\dot{n}}_{\gamma, crit}$ is confirmed at large radii in the right panel of Fig \ref{fig: w2/w2long} by extending the 
profile used above to an
outer boundary of 1500 pc in a 500 ratioed-zone simulation with $\beta$ = 1.01.

\section{Algorithm Tests: Hydrodynamics} \label{sect: franco}
\begin{figure*}
\epsscale{1.0}
\plotone{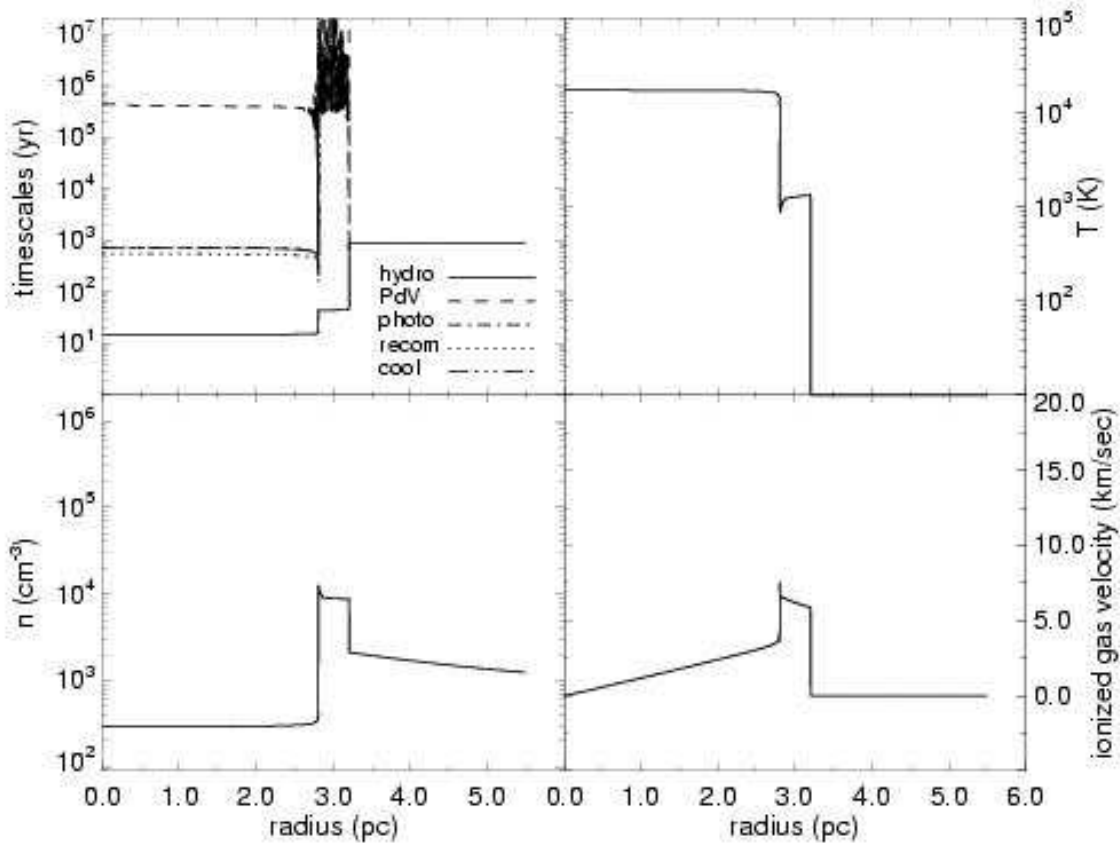}\vspace{0.15in}
\caption{Timescale, temperature, density, and velocity profiles of a D-type front in an $\omega$ = 1 density gradient
at 3.28e05 yr. \label{fig: tscl1}}
\end{figure*}

As in the \citet{ftb90} 1-D numerical tests, a source of ionizing photons with ${\dot{n}}_{\gamma}$ = 
5.0$\times$10$^{\,49}$ s$^{-1}$ was centered in a flat neutral hydrogen core with r$_{c}$ 
= 2.1$\times$10$^{\,16}$ cm, n$_c$ = 10$^{\,6}$ cm$^{-3}$, and an r$^{-\omega}$ dropoff beyond r$_{c}$.   
h$\nu$ was set to 17.0 eV and the grid was divided into 8077 uniform radial zones with 
inner and outer boundaries at 1.047$\times$10$^{\,11}$ cm and 5.50 pc, respectively.  Although 
not necessary to the benchmarks, 
for completeness we employed the same interstellar medium (ISM) cooling curve of \citet{dm72} utilized by 
\citet{ftb90} in place of the three last terms of eq. (\ref{eqn: egas}).  The photon energy and cooling curve together set the 
postfront temperature and therefore sound speed c$_{i}$ of the ionized gas.  Our choice of cooling curve and
$\epsilon_{\Gamma}$ = 2.4 eV led to postfront temperatures of 17,745 K and sound speeds of 15.6 \kms in all 
our runs, in contrast to the artificially set 10,000 K temperatures and 11.5\kms sound speeds of the 
\citet{ftb90} runs.  We applied c$_{i}$ = 15.6 \kms to the analytical expressions above for R(t) and r$_{c}$(t). The initial gas 
temperature in all the hydrodynamical tests was set to 100 K.

It should be noted that the times appearing in R(t) and r$_{c}$(t) are not taken from when the central source 
switches on.  R(t) is the position of the ionization front after the initial Stromgren sphere of radius R$_{w}$ 
has formed while r$_{c}$(t) is the location of the ionized core shock after one sound crossing time across
the core  \vspace{0.05in}
\begin{equation}
t_{c} \sim \displaystyle\frac{r_{i}}{c_{i}} \sim 1.351\times10^{\,10}\, s \vspace{0.05in}
\end{equation}
so these values must be subtracted from the total problem time when comparing the formulae to code results.   
Stromgren sphere formation times t$_{f}$ and radii emerging from the run outputs are compiled in Table 1.  In all 
cases the code computed Stromgren radii within a zone width of the predicted R$_{w}$.  
\begin{deluxetable}{ccc}
\tabletypesize{\scriptsize}
\tablecaption{Stromgren Radii R$_{w}$ and Formation Times t$_{f}$ \label{tbl-1}}
\tablehead{
\colhead{$\omega$} & \colhead{R$_{w}$ (cm)} 
                    & \colhead{t$_{f}$ (sec)}}
\startdata
  1.0   &   5.970e16    &   3.142e09  \\
  1.2   &   7.290e16    &   8.730e08  \\
  1.4   &   1.015e17    &   1.035e09  \\
  1.45  &   1.148e17    &   1.102e09  \\
  1.5   &   1.337e17    &   1.121e09  \\
\enddata
\end{deluxetable}

The derivation for R(t) does not account for the hydrodynamical details of the breakout of the shock 
through the I-front so the formula is an approximation which improves as the front grows beyond R$_{w}$.  
Similarly, r$_{c}$(t) is also an approximation which is increasingly accurate as the core shock expands beyond r$_{c}$.

\subsection{$\omega$ = 1: D-Type Front}

Column 1 of Fig \ref{fig: 123profiles} shows the density, velocity, and temperature profiles of a classic D-type ionization 
front expanding in an $\omega$ = 1 power law density for the problem times listed.  ZEUS-MP correctly predicts the formation 
of a D-type front in this gradient; the confinement of the front behind the leading shock is apparent from the abrupt 
temperature drop from 17800 K in the front to 2000 K in the shock.  The front decelerates as predicted by eq. (\ref{eqn: r_w1}), 
from 9 \kms to 6.5 \kms over the simulation time.  The dense shell thickens as the shock accumulates neutral gas, and the 
rapid decline in postfront densities as the problem evolves (especially in comparison to the initial profile) reveals the 
efficiency with which the ionized flow evacuates the cloud core.  The ionized density profile is flat because any density 
fluctuations initially present in the isothermal gas become acoustic waves that smooth the variations on timescales that 
are shorter than the expansion time of the front (which is subsonic with respect to the 17800 K gas).

\begin{figure*}
\epsscale{1.15}
\plotone{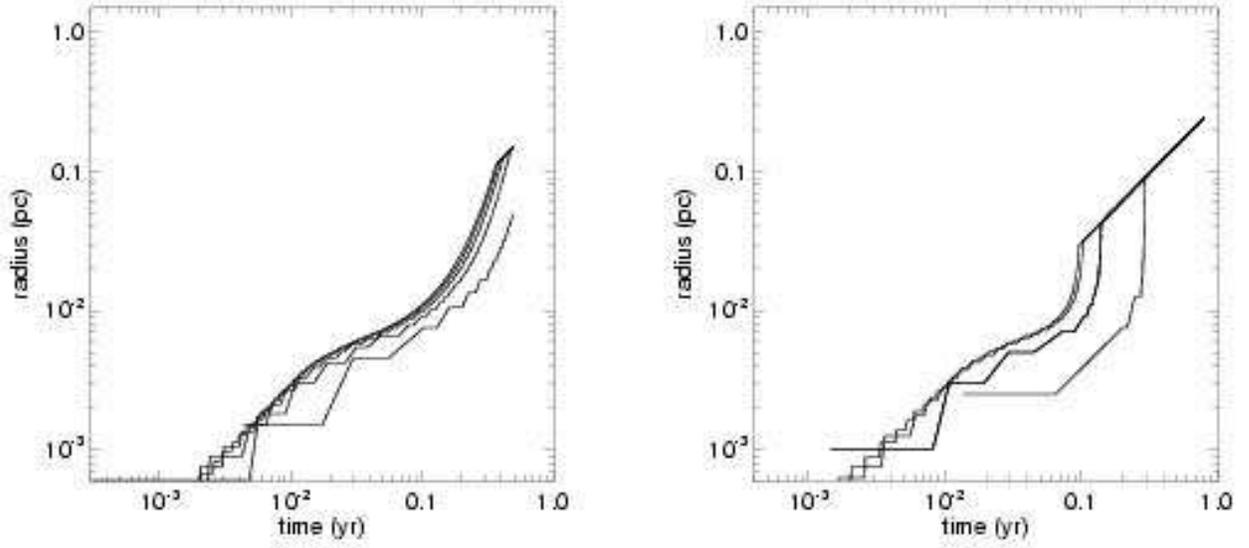}\vspace{0.15in}
\caption{Early I-front evolution simulations with 250, 500, 1000, and 2000 uniform radial zones.  Left panel: I-front 
position as a function of time in an $\omega$ = 3 gradient.  Right panel: I-front transit through an $\omega$ = 5 
gradient at the same resolutions.  \label{fig: w3front}} \vspace{0.1in}
\end{figure*}

The postfront gas remains isothermal even though central densities fall and the front performs PdV work on the shock.  This 
behavior can be understood from the various timescales governing the evolution of the gas energy density in the ionized 
flow \vspace{0.075in}
\begin{equation}
{\dot{e}}_{gas} = - p\nabla \: \cdotp \; {\bf v} \: + \: \Gamma_{phi} \: - \: \Lambda \label{eqn: eneqn}
\vspace{0.075in}  
\end{equation}
We plot photoionizational heating, PdV work, radiative cooling, and recombinational timescales defined by \vspace{0.1in}
\[ \begin{array}{c}
(\tau_{phi},\tau_{PdV},\tau_{cool},\tau_{rec}) =   \\
						   \\
\left(\displaystyle\frac{e_{gas}}{\Gamma_{phi}},\displaystyle\frac{e_{gas}}{\arrowvert 
p\nabla \:\cdotp \; {\bf v} \arrowvert},\displaystyle\frac{e_{gas}}{\Lambda},\displaystyle\frac{n_{e}}{k_{rec}n_{H^+} n_{e}}\right)   
\end{array} \vspace{0.1in}\]
%
%
%\begin{equation}
%\tau_{phi} = \displaystyle\frac{e_{gas}}{\Gamma_{phi}} \hspace{0.25in} \tau_{PdV} = \displaystyle\frac{e_{gas}}{\arrowvert 
%p\nabla \cdot \vec{\bf{v}} \arrowvert} \hspace{0.25in} \tau_{cool} = \displaystyle\frac{e_{gas}}{\Lambda}
%\vspace{0.1in}  
%\end{equation}
%

\noindent
along with the hydrodynamical Courant time in Fig \ref{fig: tscl1} as a function of radius for the $\omega$ = 1 flow.  The shortest 
timescales are associated with the most dominant terms on the right hand side of eq. (\ref{eqn: eneqn}).  The heating and cooling times 
are identical to within 0.1\% behind the D-type front, with recombination times that are slightly shorter and PdV work 
times that are three orders of magnitude greater.  Equal heating and cooling times ensure that these processes balance in 
the energy updates and do not change the gas temperature, and the much longer PdV timescales demonstrate that the work 
done by the flow behind the front is small compared to the gas energy.  Even though the ionized densities fall over time 
they still cause enough recombinations for the central star to remain energetically coupled to the flow by new 
ionizations and maintain the gas temperatures against radiative cooling and shock expansion. 

\subsection{$\omega$ = 3: R-type Front and Ionized Shock}
The early evolution of the front in the r$^{-3}$ envelope is shown in the left panel of Fig \ref{fig: w3front} with 
inner and outer problem boundaries of 1.047$\times$10$^{\,11}$ cm and 0.3 pc at resolutions of 250, 500, 1000, and 
2000 uniform radial zones.  The curves are ordered from lower right to upper left in increasing resolution.  The 
front exits the core in $\sim$ 0.05 yr through an essentially undisturbed medium since this is much shorter than the 
dynamical time of the gas.  The plots converge as the core to envelope transition becomes better resolved on the grid.  
ZEUS-MP reproduces the expected slowdown of the R-type front in the central core and its rapid acceleration as it 
exits the density gradient, having never made a transition from R-type to D-type.  Notice that beyond 0.1 pc the 
slopes of all the curves are restricted to the speed of light because the static approximation of radiative transfer 
breaks down in the rarified dropoff.  As explained in section 6, the mean free path of ionizing photons abruptly becomes
comparable to the size of the grid at that radius and an unrestricted static approximation would permit nonzero 
photoionization rates to suddenly appear all the way to the outer boundary.  As previously noted, in such circumstances 
the code restricts the position of the front to be 
\vspace{0.05in}
\begin{equation}
R_f(t) \leq c t_{problem} \label{eqn: rct}
\vspace{0.05in}  
\end{equation}
This requirement prevents superluminal I-fronts over total problem times but not necessarily over successive
timesteps, as seen in Fig \ref{fig: w3front}.  As the front crosses the core boundary R$_f$(t) briefly curves upward
with a slope greater than the speed of light c before it is abruptly limited to the speed of light by eq. (\ref{eqn: rct}).  Having been
slowed to less than c in the core the front can cross the next few zones faster than the speed of light
because because the total distance the front has traveled over the entire problem time will still be less than 
ct$_{problem}$.  This results in the slight unphysical displacement of the front upward by $\sim$ 0.1 pc,
which can be seen if one visualizes the true curve to continue up and to the right with slope c from the point where 
the computed slope becomes greater than c.  This error is negligible in comparison to the kpc scales on which the front 
later expands, and the code produces the expected approach of the front velocity to the speed of light at later times
given that the mass of the cloud is finite, as observed earlier.  In reality no cosmological density field decreases 
indefinitely so the front would eventually slow to a new Stromgren radius in the IGM \citep{wan04}. 

Columns 2 and 3 of Fig \ref{fig: 123profiles} depict the ionized flows developing in $\omega$ = 3 and $\omega$ = 5 density fields after the 
R-type ionization front has exited the cloud.  In both cases the departure of the front from the grid is visible in the 10000 K gas that
extends out to the problem boundary.  Both clouds are ionized on comparable timescales with initially flat profiles 
before much dynamic response has arisen in the gas.  However, very distinct flows emerge in the two profiles.  The 
$\omega$ = 3 field has a shallower drop with smaller pressure gradients that drive a shock that accelerates weakly but 
supersonically with respect to the outer cloud.  A reverse shock develops that can be seen in the tapered peak of the 
velocity profile and in the density maximum that falls increasingly behind the leading shock as the flow progresses.  The 
supersonic expansion of the core does not permit the central densities to relax to constant values as in the D-type front, 
but they are fairly flat out to the reverse shock because the shocked region is still dynamically coupled to the inner flow.

\begin{figure*}
\epsscale{1.0}
\plotone{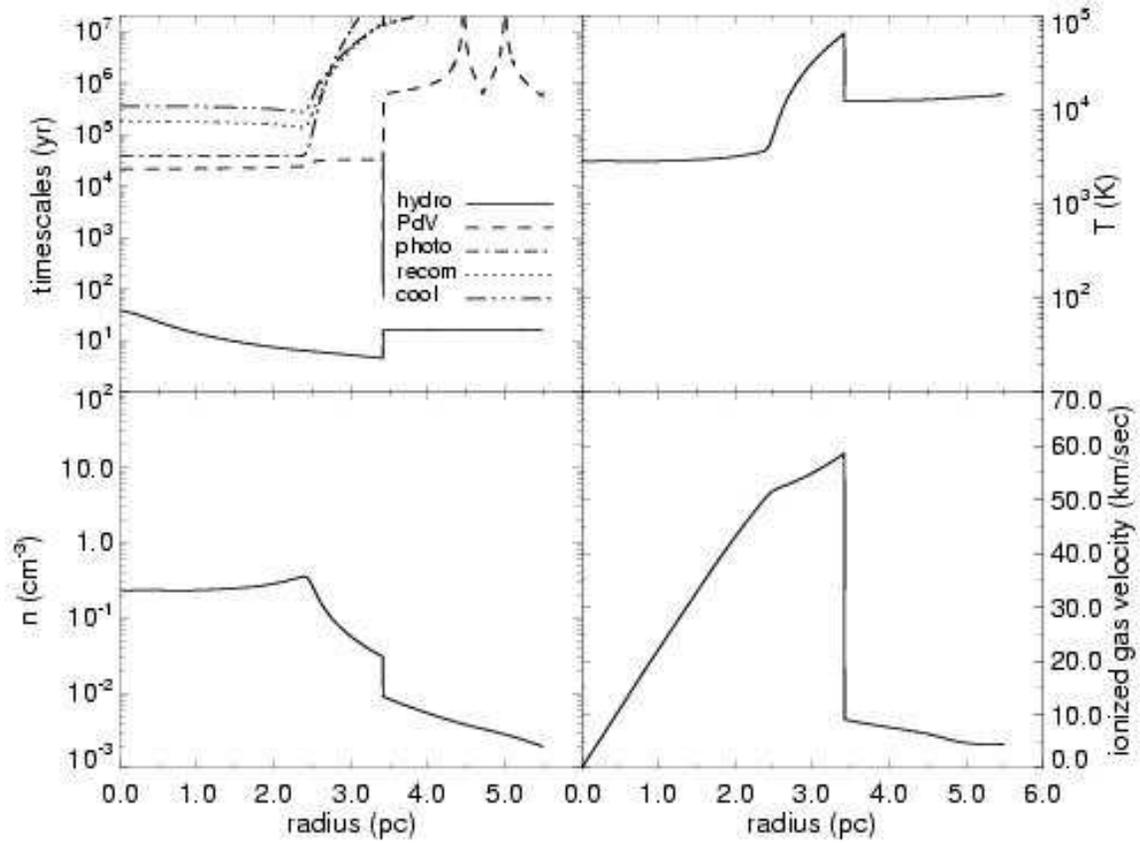}\vspace{0.15in}
\caption{Timescale, temperature, density, and velocity profiles of the ionized core shock in an $\omega$ = 3 density 
gradient at 4.41e04 yr. \label{fig: tscl2}}
\end{figure*}
\begin{figure*}
\epsscale{0.90}
\plotone{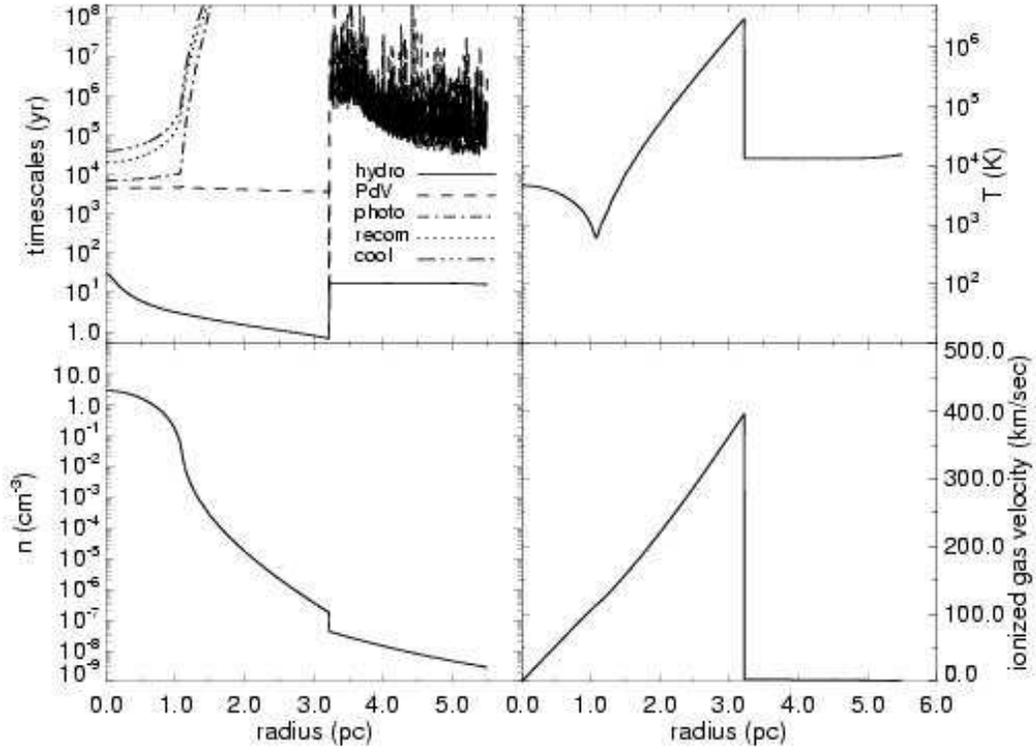}\vspace{0.15in}
\caption{Timescale, temperature, density, and velocity profiles of an ionized hypersonic shock in an $\omega$ = 5 
density gradient at 1.01e01 yr. \label{fig: tscl3}}
\end{figure*}
\begin{figure*}
\epsscale{0.99}
\plotone{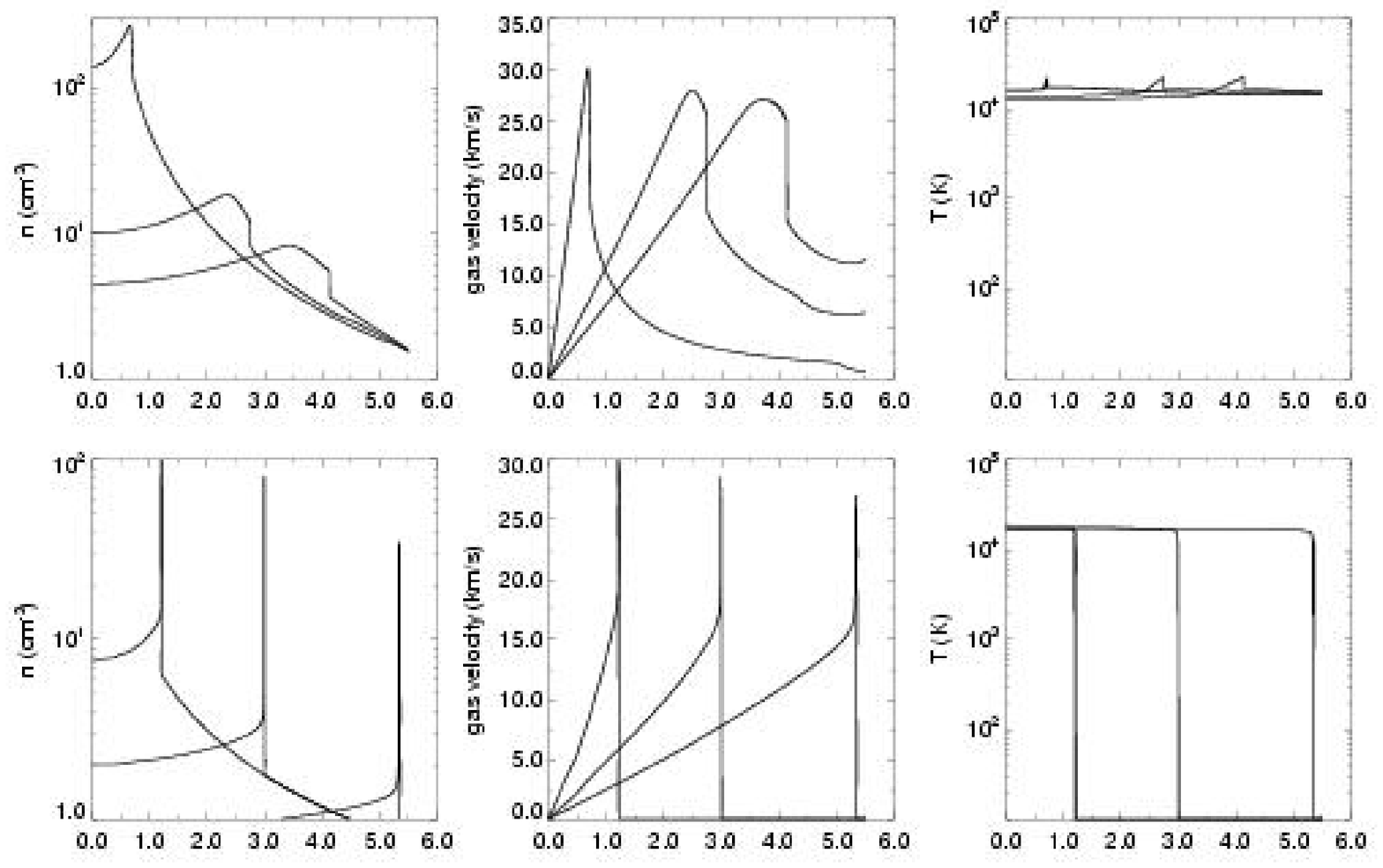}\vspace{0.15in}
\caption{Row 1 left to right: density, velocity, and temperature profiles of the ionized core shock in an $\omega$ = 2.0 
density gradient.  Output times are 1.61e04 yr, 6.04e04 yr, and 9.06e04 yr.  Row 2: density, velocity and temperature 
profiles of the D-type front in an $\omega$ = 1.42 radial density.  Output times are 3.67e04 yr, 9.79e04 yr, and 1.83e05 
yr. \label{fig: 215profiles}}
\end{figure*}

At any given moment the postshock gas temperatures are uniform in radius up to the reverse shock, but they decrease over 
time as the flow is driven outward.  The temperatures are almost flat because the gas heating and cooling rates are nearly 
constant in the fairly uniform central densities.  The temperatures fall with time because the central gas densities are 
lower than in the $\omega$ = 1 flow above.  There are fewer of the recombinations that enable the central star to sustain the 
postshock temperatures as in the D-type front, so the flow expands partly at the expense of its own internal energy.  The 
timescales of eq. (\ref{eqn: eneqn}) governing the gas internal energy and the Courant times are shown in Fig \ref{fig: tscl2}.  PdV work done by the gas 
clearly outpaces photoionizational heating out to the reverse shock with a net loss of internal energy in the ionized gas.  
The suppression of recombinations is evident in the recombination times, which are nearly ten times the photoheating 
times. 

\subsection{$\omega$ = 5: Hypersonic Ionized Flow}

We show the initial transit of the I-front through an r$^{-5}$ falloff in the right panel of Fig \ref{fig: w3front} for the 
same boundaries and grid resolutions as in the $\omega$ = 3 plots on the left; the curves again progress from lower 
right to upper left as the resolution increases.  ZEUS-MP again correctly predicts the initial slowing of the front in the
core and its rapid acceleration down the gradient without ever becoming D-type.  There are no qualitative differences 
between the fronts in the two gradients prior to their exit from the core because their initial conditions are identical:
the $\omega$ = 5 front also crosses the core in $\sim$ 0.05 yr.  There is likewise little difference between the two fronts
after their velocities have been limited to the speed of light by the code.  The two prominent differences are the slower 
convergence of the curves beyond r$_c$ in the $\omega$ = 5 gradient and their greater superluminal velocities before being
set to c by the restriction algorithm.  Both trends are expected: the front will descend the steeper gradient more quickly
before being limited to the speed of light and more grid cells are required to resolve the falloff.  Again, because the 
$\omega$ = 5 cloud has finite mass, the velocity of the front becomes c at later times.

In this profile a much more abrupt isothermal density dropoff remains in the wake of the R-type front.  Extreme pressure gradients 
launch the edge of the core outward in a strong shock that is essentially a free expansion, as seen in the velocity profiles 
in column 3 of Fig \ref{fig: 123profiles}.  The density falls off so sharply that recombinations are quenched there and the shock expands 
adiabatically.  The shock advances so quickly that it becomes dynamically decoupled from the flow behind it, so no reverse shock 
forms and all the flow variables remain stratified with radius.  Temperatures rise to 4 $\times$ 10$^{6}$ K in the shock but 
drop to 500 K behind it from adiabatic expansion.  The sequence of velocity profiles confirms that the core shock accelerates 
over the entire evolution time of the problem, as predicted by eq. (\ref{eqn: om5}).  It becomes hypersonic with speeds in excess of 
400 \kms.  The central star is least energetically coupled to this ionized flow, with much of it being driven by its own 
internal energy. 

Temperatures in the central ionized gas fall with time for the same reasons as in the $\omega$ = 3 profiles.  The energy and 
Courant times for the $\omega$ = 5 flow appear in Fig \ref{fig: tscl3}.  The hierarchy of timescales is similar to that of the $\omega$ = 3 
flow up to the position of the shock, at which point cooling, heating, and recombination times rise much more quickly than 
their $\omega$ = 3 counterparts.  The steeper jump in timescales is due to the strong suppression of recombinations in the 
highly stratified densities.

\subsection{$\omega$ = 1.5 and 2.0}

ZEUS-MP accurately reproduces the mildly-accelerating ionized core shock expected to form in r$^{-3}$ radial densities as 
well as the strong adiabatic shock predicted for r$^{-5}$ gradients, with hydrodynamical profiles that are in excellent 
agreement with earlier work performed by 1-D implicit lagrangian codes \citep{ftb90}.  In this section we summarize our 
code results for I-fronts and shocks in r$^{-1.5}$ and r$^{-2.0}$ core envelopes and present their density, velocity, and 
temperature profiles in Fig \ref{fig: 215profiles}.  ZEUS-MP verifies the theoretical prediction that a constant-velocity ionized shock forms 
in the $\omega$ = 2 gradient after the rapid departure of the R-type front, as seen in the upper middle panel of Fig \ref{fig: 215profiles}.  This 
ionized flow does not steepen into as strong a shock as in the $\omega$ = 3 case, which is evident from the lower gas 
velocities and smaller spike in the temperature distribution.  Timescale analysis indicates that heating and recombination 
times are nearly equal in the postshock gas but not to the same degree as in the D-type $\omega$ = 1 front, so expansion 
occurs partly at the expense of the internal energy of the gas.  Consequently, the level gas temperatures behind the shock 
decrease slightly over time but not to the extent in the $\omega$ = 3 shock.  The evolution of the postshock gas is 
intermediate between that in the r$^{-1}$ and r$^{-3}$ cases.  This flow regime is relevant to primordial minihalos: high
resolution simulations of the formation of these objects yield spherically-averaged baryon density profiles with 2.0 $<$
$\omega$ $<$ 2.5 \citep{abn02}.  The correspondence is not exact because $\omega$ can vary in radius over this range, which
is why both shock accelerations and decelerations are observed numerically in profiles derived from cosmological initial
conditions \citep{wan04}.

Our code confirms that the $\omega$ = 1.5 front evolves essentially as D-type but with no shocked neutral shell: the front 
and the shock are coincident and advance at the same velocity.  Throughout the evolution of this flow the front is precariously balanced at 
the edge of breakout through the shock and down the gradient.  This breakthrough is sensitive to small errors in shock 
position at early times; as a result, ZEUS-MP finds that the I-front overruns the shock when $\omega$ = 1.45 instead of 1.5 
(the densities, velocities, and temperatures in the second row of Fig \ref{fig: 215profiles} are for an r$^{-1.42}$ distribution).  However,
this is a relatively small error, and our algorithm captures shock and front positions to within 10\% of theory for $\omega$ 
= 1.4 and 1.6.  Although this discrepancy is still under investigation, we believe it to be due to small errors in energy
conservation in our eulerian code.  In this respect lagrangian codes would likely enjoy an advantage in
proper shock placement because of their conservative formalism.

\subsection{Hydrodynamical Convergence}

\begin{figure*}
\epsscale{1.2}
\plotone{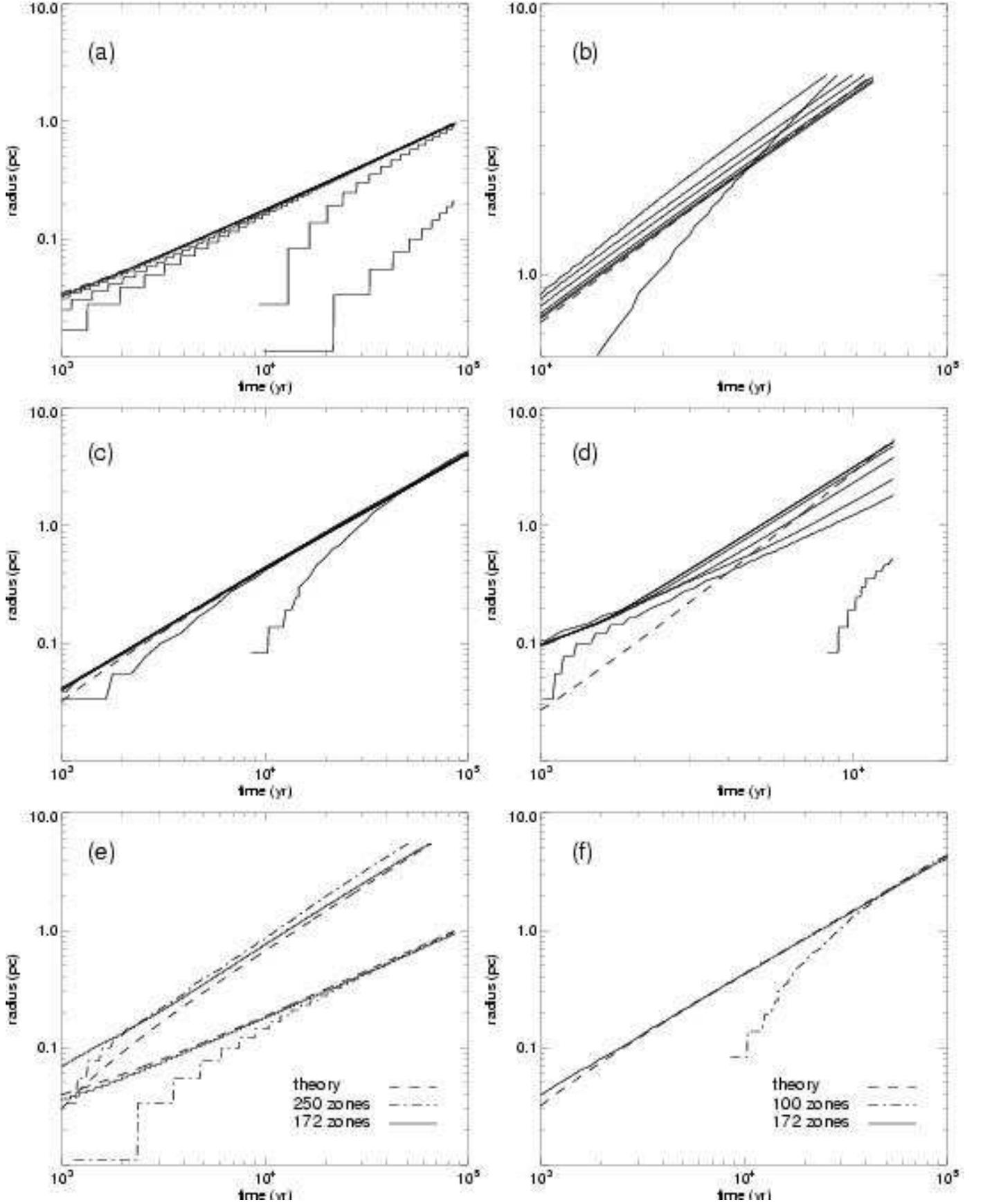}\vspace{0.15in}
\caption{Upper and middle rows: convergence simulations with 100, 250, 500, 1000, 2000, 4000, and 8000 uniform radial 
zones.  Panel (a): D-type I-front position as a function of time in an $\omega$ = 1 radial gradient. Panel (b): time 
evolution of a weakly accelerating $\omega$ = 3 ionized shock. Panel (c): position of a constant-velocity ionized core 
shock in an $\omega$ = 2 density gradient with respect to time. Panel (d): a strongly accelerating hypersonic core shock 
in an $\omega$ = 5 dropoff.  Third row: runs with both uniform and ratioed zones compared to theory. Panel (e): $\omega$ 
= 1 and 3 runs with 250 uniform zones and 172 ratioed zones. Panel (f): $\omega$ = 2 runs with 100 uniform zones and 172 
ratioed zones.\label{fig: shkres}}
\end{figure*}
We demonstrate the convergence of our algorithm to eq. (\ref{eqn: r_w1}) for the I-front position R$_\omega$(t) 
in an $\omega$ = 1 dropoff as well as to the ionized shock position eqs. (\ref{eqn: om2}), (\ref{eqn: om3}), and (\ref{eqn: om5}) for $\omega$ = 2, 3 and 5 gradients in the first two rows of Fig \ref{fig: shkres}.  For 
each density regime we utilized grids with 100, 250, 500, 1000, 2000, 4000, and 8000 uniform radial zones.  The dashed line
in each panel of Fig \ref{fig: shkres} is the analytical solution.  In Fig \ref{fig: shkres}a the numerical
predictions for the position of the $\omega$ = 1 D-type ionization front are ordered from lowest to highest resolution from the lower 
right corner upward to the left toward the expected solution.  ZEUS-MP converges to within 10\% of eq. (\ref{eqn: r_w1}) by 
2000 zones and to within 2 - 3\% at 8000 zones (the 4000 zone curve is indistinguishable from the 8000 zone solution at
this scale).  Because eq. (\ref{eqn: r_w1}) does not account for the R- to D-type transition of the front, the agreement 
improves the further the front advances beyond the Stromgren radius at any resolution.

Convergence was fastest in the case of the constant velocity ionized core shock in the $\omega$ = 2 gradient, as shown in 
Fig \ref{fig: shkres}c.  The order of solutions is again from lowest to highest resolution from the lower
right upward to the left toward eq. (\ref{eqn: om2}).  Agreement to within 10\% between 10$^3$ yr and 10$^4$ yr and 2 - 3\% 
past 10$^4$ yr was achieved with just 500 zones.  The trend is somewhat different for the weakly accelerating ionized shock
in the $\omega$ = 3 panel in Fig \ref{fig: shkres}b.  The numerical models are all somewhat above the analytical curve at later 
times but the 100 and 250 zone runs are below it at earlier times.  The different grids are most easily distinguished by 
their overshoot at late times: with the exception of the 100 zone curve they converge downward toward eq. (\ref{eqn: om3})
with increasing resolution, with the 8000 zone curve dipping slightly below it near the end of the evolution time.  With 
2000 zones the convergence is to within 10\% of the expected values between 5000 yr and 10 kyr and to within 1.5\% by 60 
kyr.

In the highly supersonic $\omega$ = 5 core shock runs shown in Fig \ref{fig: shkres}d, the progression
of the numerical curves toward eq. (\ref{eqn: om5}) is again from lower right to upper left as the resolution increases,
but they converge to a solution that lies well above the analytical result.  Although the hydrodynamical profiles for the
$\omega$ = 5 ionized flow in Fig \ref{fig: 123profiles} are in excellent qualitative agreement with earlier work 
\citep{ftb90}, in this gradient the algorithm does not accurately compute the shock placement for reasons that are currently under investigation.
This flow regime is rather extreme, with very high shock temperatures that challenge the energy conservation of our code.  Such 
flows are unlikely to develop in the 3-D photoevaporation of the high-redshift minihalos that we plan to study.  In the 2 $<$ 
$\omega$ $<$ 3 gradients of relevance to those primordial structures \citep{abn02}, ZEUS-MP is quite accurate.  

The convergence trends in the early I-front evolution studies in the past three sections suggest that a key factor in 
accurate front placement is proper resolution of the central core and its initial dropoff.  Ratioed grids with higher 
central and lower outer resolutions can capture proper shock placement and front positions with far fewer zones 
than uniform grids, as shown in the bottom row of Fig \ref{fig: shkres}.  In Fig \ref{fig: shkres}e we plot ionized
core shock radii as a function of time in both $\omega$ = 1 and 3 density profiles for a 250 zone uniform grid and a 172
zone ratioed grid with $\beta$ = 1.03 as defined in section \ref{sect: rmw static}; the analytical prediction in both 
panels is again the dashed line.  We show the corresponding simulations for an $\omega$ = 2 profile in Fig \ref{fig: shkres}f but with a 100 zone 
uniform grid.  We achieve the same accuracy with 172 ratioed zones as with 2000 uniform zones in the first two regimes 
and as with 500 zones in the $\omega$ = 2 gradient (the most convergent of the four regimes tested).  Equal accuracy 
with large savings in computational resources motivated our use of ratioed grids in earlier work, with the only 
sacrifice being the loss of some detail in the hydrodynamical profiles (e.g. compare Fig 3 in \citet{wan04} to the 
$\omega$ = 2 profiles in Fig \ref{fig: 215profiles}).  

\section{Timestep Evolution}

The physical timescales governing code timesteps as I-fronts and flows evolve on the grid yield insights about the 
processes that drive the fronts and are the key to future algorithm optimization.  In general, different timescales 
dominate the transit of the front than the rise of ionized flows behind it.  We examine how these timescales control 
the advance of the solution in four density gradients:  $\omega$ = 1.0, 2.0, 3.0, and 5.0.  As a rule, the photoheating 
timescales in the cell being ionized while the front remains on the grid determine the global timestep over which the 
entire solution is updated.  We first describe how cells approach ionization equilibrium at different radii in these 
gradients.  

\begin{figure*}
\epsscale{1.2}
\plotone{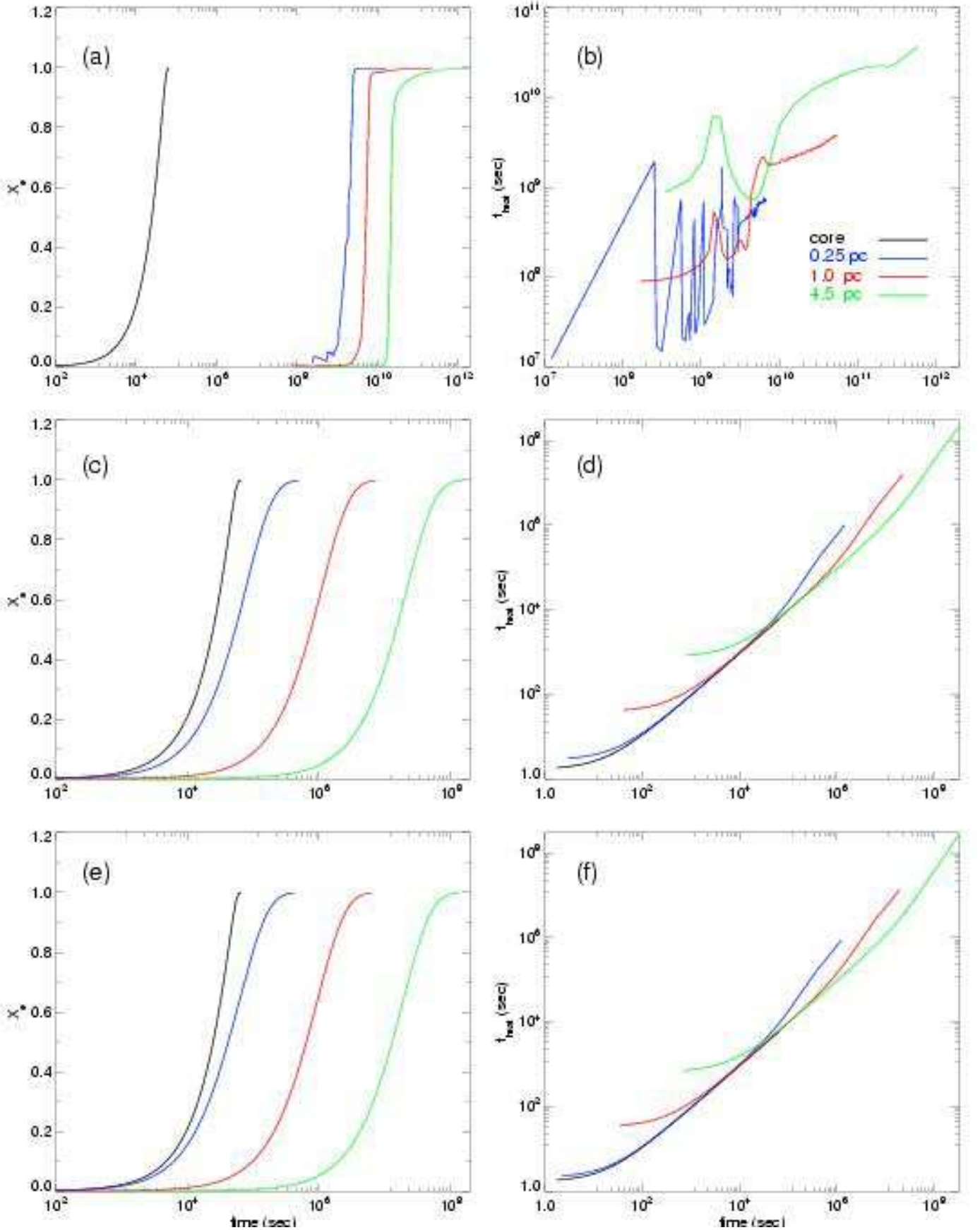}
\caption{Ionization fraction and t$_{heat}$ as a function of time in zones in the core and at r = 0.25 pc, 1.0 pc, 
and 4.5 pc as they approach ionization equilibrium.  Panels (a), (c), and (e):  ionization fractions in $\omega$ = 
1, 3, and 5, respectively.  Panels (b), (d), and (f): the evolution of t$_{heat}$ as a function of time in $\omega$ 
= 1, 3, and 5, respectively. \label{fig: ionprfl}}
\end{figure*}

\subsection{Ionization Equilibrium of a Single Zone}

The ionization fraction X$_e$ of a zone as a function of time varies with distance from the central source, ambient density, 
and type of ionization front.  In this section we first examine the equilibration of a zone in the flat central core (the 
same in all four density gradients) and then analyze three other zones at radii of 0.25 pc, 1.0 pc, and 4.5 pc, well into 
the density falloffs in which the distinct characters of the fronts emerge.  We plot the ionization curves for the four 
radii in Figs 
\ref{fig: ionprfl}a, c, e, and Fig \ref{fig: tsclprfl}a for $\omega$ = 1, 3, 5, and 2 gradients, respectively.  We tally 
the corresponding consecutive heating timesteps t$_{heat}$ (eq. (\ref{eqn: theat})) in the four cells as they come to 
equilibrium in Figs \ref{fig: ionprfl}b, d, f, and Fig \ref{fig: tsclprfl}b (recall that t$_{heat}$ is the interval over 
which full hydrodynamical updates are performed).

\subsubsection{Central Zones}

The zone we consider in the central core is the fourth from the coordinate origin, but the ionization of the inner boundary 
also merits discussion.  Chemical subcycling over t$_{heat}$ is uniquely extreme in this zone because the UV radiation 
entering its lower face initially encounters no electron fraction whatsoever.  Since n$_e$ is zero t$_{chem}$ is extremely 
small so far more chemical timesteps are invoked to cover t$_{heat}$ than are required by the network (several thousand).  
In practice this only occurs on the inner boundary because over the 80 - 90 heating times required to ionize the inner zone 
the hydrodynamic updates advect a small electron fraction into the next zone, substantially reducing the number of initial 
chemical cycles there.  The network subcycles $\sim$ 100 times in the fourth zone in its first heating timestep but only 
two or three times by the fifth timestep.  Roughly 100 t$_{heat}$ are necessary to ionize this zone.

As shown in Fig \ref{fig: ionprfl}d, heating times are at first $\sim$ 1 s in the central zone but
increase to 1 $\times$ 10$^{4}$ s by the time it reaches X$_e$ = 0.999.  The heating rate ${\dot{e}}_{rad}$ is greatest when 
the front enters the cell because k$_{ph}$n$_H$$\epsilon_{\Gamma}$ is at its maximum but cooling rates are at their minimum
because of the low initial temperature of the cell.  This low temperature implies that e$_{gas}$ is also at its minimum,
ensuring that the first t$_{heat}$ is the smallest.  As the temperature and gas energy rise with X$_e$, cooling processes are
activated and decrease ${\dot{e}}_{rad}$, increasing t$_{heat}$.  When X$_e$ exceeds 0.999 heating times can abruptly dwarf the 
light-crossing time t$_{lc}$ of the cell because cooling suddenly balances photoionizational heating, driving ${\dot{e}}_{rad}$ 
sharply downward.  The code solution consequently takes too large a timestep forward, over which ionization that should have 
commenced in subsequent zones cannot, unphysically stalling the young I-front.  This only occurs in core zones close to the 
source because t$_{ioniz}$ $\gg$ t$_{lc}$ further from the center.  Ionization begins in the next zone before the previous zone 
can come to equilibrium, decreasing the code timestep dt and preventing anomalous jumps of the solution forward in time.  The 
jumps in central zones are easily remedied by limiting t$_{heat}$ to be less than a tenth of the light-crossing time there. 
After a core zone is fully ionized the code solution advances by consecutive 0.1 t$_{lc}$ steps until eq. (\ref{eqn: rct}) permits 
ionizations in the next zone.  Hence, the UV flux in the core zones is a step function in time, immediately rising to full 
intensity because intervening zones are completely transparent. 

\subsubsection{$\omega$ = 3 and 5}

The heating time profiles for R-type fronts in $\omega$ = 3 and 5 gradients appear in Figs \ref{fig: ionprfl}d and f.  As noted earlier, these fronts rapidly accelerate and must be limited to the speed of light by the code.  
The UV flux in the outer zones is again a step function in time because, as observed above, the front arrives at the next zone 
before the current zone is completely ionized.  In this instance the step in the radiation field is not quite to its full intensity 
because one or more of the intervening zones is not fully transparent.  As in the core zone, we begin the tally of t$_{heat}$ 
in the outer three test zones when their photoionization rates switch on.  The heating times are again comparatively small at 
first and grow by several orders of magnitude as the cells become ionized.  The heating times as a rule increase with distance 
from the source because the smaller outer fluxes generate lower photoionization rates.  

The static approximation is clearly violated near the inner boundary because the 1 $\times$ 10$^{4}$ s ionization timescale of 
the core zone is comparable to its 7 $\times$ 10$^{4}$ s light-crossing time.  The approximation also fails in the outer zones 
of these two gradients even though ionization times are much longer than light-crossing times there.  The explanation lies in
the photon mean free paths:  as shown in Table 2, they exceed the length of the grid as the densities plummet with radius.  If 
the widths of the fronts overrun the outer boundary the static approximation would allow nonzero ionization rates in zones 
that could not have been reached by light by that time in the simulation.  We must again employ eq. (\ref{eqn: rct}) to prevent 
superluminal velocities in the outer regions.  IGM mean densities prevent runaway I-fronts in realistic cosmological conditions.
In compiling heating time profiles of the outer zones we terminate the advance of the front beyond the zone in question to 
avoid the downshift in code timestep associated with ionizations in a new zone before the current cell has equilibrated.  

\begin{deluxetable}{ccccc}
\tabletypesize{\scriptsize}
\tablecaption{UV photon mean free paths.\label{tbl-2}}
\tablehead{
\colhead{radius (pc)} & \colhead{$\omega$ = 1 } 
                    & \colhead{$\omega$ = 2} & \colhead{$\omega$ = 3} & \colhead{$\omega$ = 5}}
\startdata 
 0.25 & 6.35e-6 pc & 1.29e-4 pc & 4.79e-3 pc &   6.6 pc      \\
 1.0  & 1.38e-5 pc & 2.03e-3 pc & 0.299 pc   &   5600 pc     \\
 4.5  & 6.21e-5 pc & 4.10e-2 pc & 27.14 pc   &   1.18e07 pc  \\
\enddata
\end{deluxetable}

\subsubsection{$\omega$ = 2}

\begin{figure*}
\epsscale{1.2}
\plotone{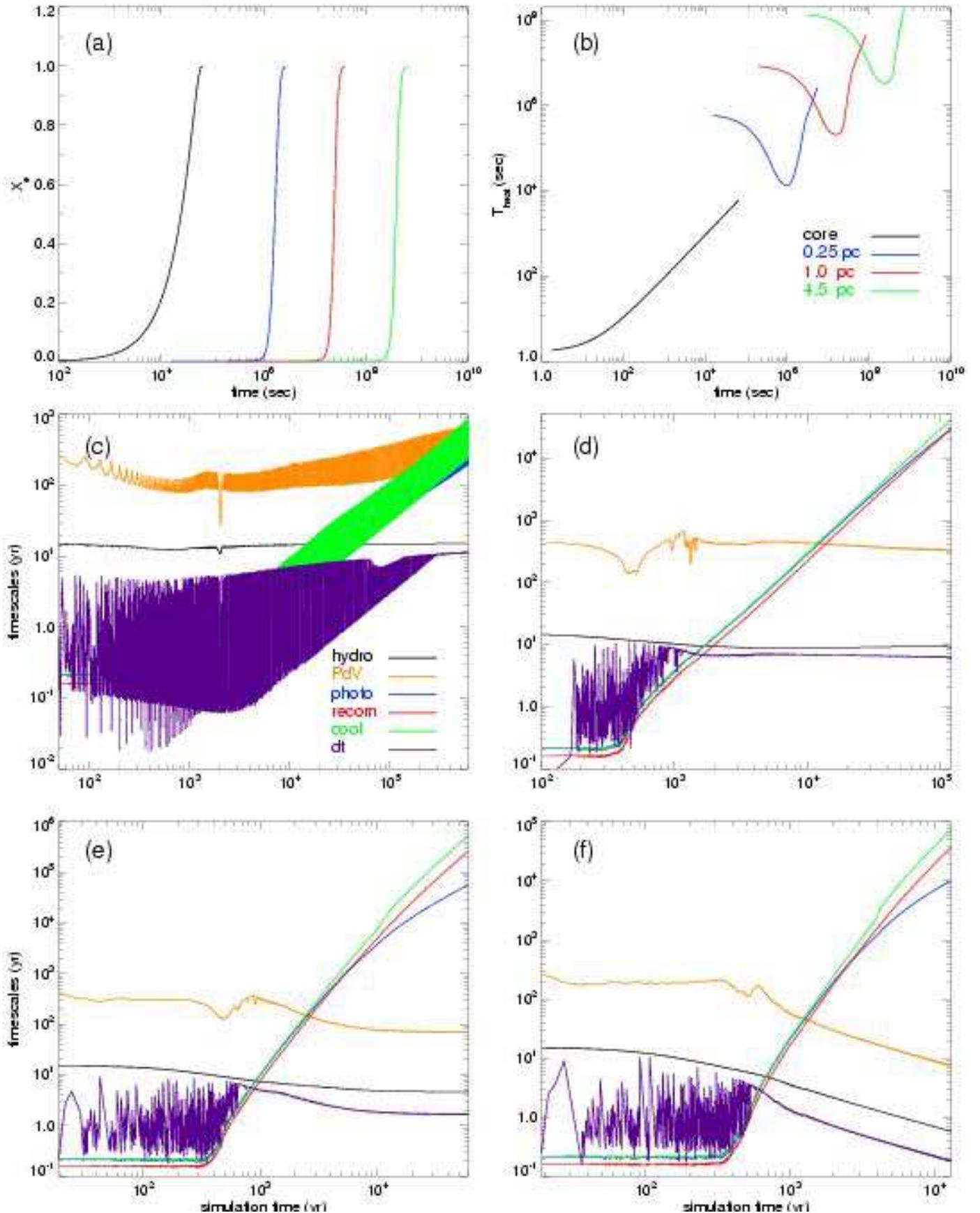}
\caption{Row 1: same as in Fig \ref{fig: ionprfl} but for an $\omega$ = 2 density gradient.  Panel (a): ionization fraction.
Panel (b): heating/cooling time t$_{heat}$.  Rows 2 and 3: global minima of the Courant, PdV work, photoionizational heating, 
recombination, and cooling timescales along with code timestep dt as a function of time after the fronts have exited the grids 
(except in the $\omega$ = 1 case).  Panels (c) and (d): $\omega$ = 1 and 2 density gradients, respectively.  Panels (e) and 
(f):  $\omega$ = 3 and 5 density profiles, respectively.
\label{fig: tsclprfl}}
\end{figure*}

In Fig \ref{fig: tsclprfl}b the $\omega$ = 2 t$_{heat}$ profiles at 0.25 pc, 1.0 pc, and 4.5 pc have a more complicated structure because the UV 
radiation illuminating those zones is not a step function in time.  As seen in Table 2 the front is a fraction of a zone in width 
at first but quickly widens to encompass several zones.  Although R-type, the front has a velocity well below the speed of light 
so ionizations proceed several zones in advance of the center of the front.  The radiation field at the leading zone is initially 
weak (having been attenuated by the previous few partially-ionized zones) but soon grows to its full intensity as the center of 
the front crosses the cell.  As a result the energy deposition into the gas is first relatively small but increases as the 
radiation field intensifies.  ${\dot{e}}_{rad}$ then dips back downward when cooling processes are switched on as the zone 
temperature increases.  The heating times therefore curve downward but then recover upward as shown in Fig \ref{fig: tsclprfl}b.  The profiles again migrate upward with distance from the source.  

The static approximation is well-obeyed in this regime given that the ionization times are much longer than the zone-crossing 
times for light and that the speed of the front is much smaller than c.  The width of the front also remains small in comparison 
to the grid.  We begin the heating time tally in the three cells when X$_e$ has risen to 1 $\times$ 10$^{-5}$ because there is no 
sudden activation of photoionization rates in them. 

\subsubsection{$\omega$ = 1}

As we show in Figs \ref{fig: ionprfl}a and b, test zones ionized by the D-type front in an $\omega$ = 1 gradient exhibit 
noticeably different ionization fractions and heating times because the shock reaches them before the front.  The shock raises 
their temperatures to a few thousand K and induces collisional ionizations that leave a residual X$_e \sim$ 0.01 - 02.  Shock
heating lengthens t$_{heat}$ when the front reaches the cell because e$_{gas}$ is 200 - 300 times greater than in the preshock gas 
and the initial chemistry subcycling is heavily reduced by the collisional electron fraction.  The radiation profile in these cells 
is a smooth function of time because of the UV photons escaping ahead of the front into the slightly preionized zone (the width 
of the front itself in these densities is less than a tenth of a zone).  We initiate the tally of t$_{heat}$ in the cell when 
ionization rates surge with the arrival of the center of the front.

The large fluctuations in heating times at 0.25 pc are due to postshock numerical oscillations of the hydrodynamical flow 
variables.  These variations dampen at larger radii because the shocked neutral shell thickens with time; the densities the
front photoevaporates have more time to numerically relax to steadier values.  Postshock oscillation is also responsible for 
the early variations in the 0.25 pc X$_e$ profile.  As in the other gradients, t$_{heat}$ again increases with distance from 
the central source.  The static approximation is most valid in this density regime, with material properties changing many 
orders of magnitude more slowly than light-crossing times.

\subsection{Global Timestep Control: R-type/D-type Fronts}

As the nascent I-front emerges from the central core the numerical solution evolves in a succession of heating timesteps that 
rise and then fall as one zone comes to equilibrium and ionization commences in the next.  As discussed earlier, the heating
timesteps in the zone being ionized rarely achieve their full values in Figs \ref{fig: ionprfl} and \ref{fig: tsclprfl}
because ionization in the next zone is activated before the current zone can come to full equilibrium.  Because the timesteps 
are determined by the sum of photoionizational heating and cooling in the zone being ionized, they are much smaller than Courant
times in the postfront gas ($\sim$ 10 yr) while the front is on the grid.  Although restrictive, such timesteps are necessary 
in part because of the proximity of the ionized gas temperature to the sharp drop in the cooling curve near 10,000 K; longer 
timesteps can cause sudden unphysical cooling in equilibrated zones that lose energy in a heating cycle.  Whenever ionization 
fronts approach the speed of light the solution is limited to relatively short steps until the front exits the grid (although 
it should be remembered from Figs \ref{fig: ionprfl} and \ref{fig: tsclprfl} that the timesteps grow as the radius of the front 
increases).  The rise and sharp fall of timesteps continues until the I-front exits the outer boundary.  In $\omega$ = 3 and 5 
gradients the code expends 85\% of its cycles propagating the front across the grid ($\sim$ 19 yr) and its last 15\% advancing 
the core shock to the outer boundary ($\sim$ 1 $\times$ 10$^4$ yr).  

The transport of R-type fronts in $\omega$ = 2 gradients is uniquely efficient because of the unusual t$_{heat}$ profile of 
the zones.  When a cell begins to be ionized the heating times are initially fairly long and are again relatively long as the 
zone approaches equilibrium.  On average, this curve allows the solution to evolve much further in time each cycle.  70\% of 
the simulation timesteps are utilized for the transit of the front ($\sim$ 500 yr) with the remainder being spent on the exit 
of the ionized shock ($\sim$ 1.5 $\times$ 10$^5$ yr).  This simulation requires only 20\% of the CPU time of the $\omega$ = 3 
and 5 runs, which require about the same time to execute.  The much longer heating times in the $\omega$ = 1 gradient enable
the code to advance the D-type front across the grid with only 30\% of the cycles needed in the $\omega$ = 3 and 5 models.  
However, much longer simulation times ($\sim$ 6.0 $\times$ 10$^5$ yr) are necessary to advance the front to the outer boundary 
because of its relatively low velocity behind the shock, so this run executes with somewhat greater CPU times than the $\omega$ 
= 2 model.
  
\subsection{Global Timestep Control: Ionized Flows}

In Figs \ref{fig: tsclprfl}c, d, e, and f we plot global minima of the Courant, PdV work, photoionizational heating, 
recombination, and cooling timescales along with the code timestep dt as a function of time after the fronts have exited 
the grid (except for the $\omega$ = 1 case) for $\omega$ = 1, 2, 3, and 5 density profiles.  After the outermost zone is 
completely ionized the minimum heating and 
cooling times by which the solution is governed settle to nearly constant values for the sound crossing time of the central 
core, approximately 500 yr.  The 1 $\times$ 10$^7$ - 10$^8$ s code timesteps dt are similar to the heating times in core 
zones that have come to complete equilibrium (see Fig \ref{fig: ionprfl}).  500 yr marks the steepening of the pressure wave 
at the core's edge into the ionized shock that begins to evacuate the core.  As central densities fall recombinations are
supressed, slowing cooling rates and new ionizations.  Photoheating, recombination, and cooling timescales rise, surpassing 
Courant times at t $\sim$ 1000 yr, at which point the code adopts t$_{Cour}$ as the new global timestep.  The noise in the 
code timestep prior to crossover is due the nearly equal heating and cooling rates; their difference is sensitive to minor 
variations in each rate and can modulate relatively large fluctuations in $e_{gas}/{\dot{e}}_{rad}$.  This noise disappears 
after the algorithm turns to t$_{Cour}$ to compute dt.  The minimum t$_{photo}$,t$_{recom}$, and t$_{cool}$ originate in the 
relatively flat densities behind the shock (see Fig \ref{fig: tscl2}) and continue to rise with the expulsion of baryons 
from the center of the cloud.  The three timescales branch away from each other at late times in the $\omega$ = 3 and 5 panels 
because the cooling rates are sensitive to the drop in postshock gas temperature (from 18,000 K down to 2000 K in the $\omega$ 
= 3 gradient) as the ionized flows expand in the course of the simulation.  The rates do not diverge in $\omega$ = 1 and 2 
postshock flow because the gas is either isothermal or nearly so.

The Courant time is approximately 15 yr in the newly ionized gas in all four cases but begins to fall after the core sound 
crossing time in the last three regimes as a result of the formation of the shock.  The minimum Courant time coincides with 
the temperature peak of the shock at all times thereafter in the $\omega$ = 2, 3, and 5 flows but resides within the I-front 
in the $\omega$ = 1 gradient because the much higher temperature of the ionized gas in comparison to the weakly shocked neutral 
gas.  As shown in Fig \ref{fig: tsclprfl}c, the Courant profile is almost flat in this flow regime because 
the postfront gas is isothermal.  The ionized core shock in the $\omega$ = 2 gradient has a nearly constant velocity after 
formation so its Courant time remains flat after falling to 10 yr in Fig \ref{fig: tsclprfl}d.  The $\omega$ = 3 flow only weakly accelerates and therefore 
exhibits a similar profile in Fig \ref{fig: tsclprfl}e with final Courant times of 4.5 yr in the stronger shock.  Courant times in the hypersonic $\omega$ 
= 5 flow continue to fall in Fig \ref{fig: tsclprfl}f because of its strong acceleration, eventually dropping to 0.6 yr by the end of the simulation.

The first dip in the PdV timescale at t $\sim$ 480 yr in the $\omega$ = 2, 3, and 5 plots results is due to outflow of heated 
gas from the inner boundary zone.  The oscillations after the dip at t $\sim$ 800 yr are associated with the launch of the core 
shock.  The minimum work times later track the forward edge of the shock at all times thereafter in all four regimes.  The 
behavior of the $\omega$ = 2, 3, and 5 PdV profiles after the rise of the shock mirrors that of the Courant profiles.  The 
constant speed $\omega$ = 2 shock uniformly accelerates parcels of upstream gas throughout its evolution so its PdV timescales 
remain fairly flat.  In contrast, the strong $\omega$ = 5 shock accelerates upstream fluid elements to increasingly higher 
speeds as it advances so its work timescales continue to fall.  The profile of the gently accelerating $\omega$ = 3 shock falls 
in between these two cases, sloping downward and then evening out at later times.  PdV timescales in the D-type front actually 
rise over time because of the deceleration of the shock.  The momentary spike in the profile at t $\sim$ 2000 yr (also manifest 
in the Courant profile) occurs at the inner boundary and is likely due to pressure fluctuations associated with acoustic waves 
there.

Several features in the $\omega$ = 1 plot deserve attention.  The PdV timescale oscillations at t $<$ 400 yr are from sound
waves in the flat core.  The postshock numerical oscillations of the flow variables responsible for the rapid fluctuations in the
heating profiles continue to cause rapid variations in the heating, cooling, and work timescales throughout the simulation, 
dominating the code timestep until the average t$_{heat}$ rises above the Courant time at t $\sim$ 3 $\times$ 10$^5$ yr.  It is
important to remember that the noise in t$_{heat}$ and t$_{cool}$ in this panel is not readily apparent in Fig \ref{fig: tscl1}
because the global minima in Fig \ref{fig: tsclprfl} occur at the rear edge of the shock.  The two timescales quickly become 
equal behind the shock.

\section{The Role of Radiation Pressure in I-front Transport}

%  
%%\begin{figure*}
%%\epsscale{0.65}
%%\plotone{f15.eps}\vspace{0.15in}
%%\caption{Thermal pressure gradient, radiation pressure, and gravitational accelerations taken at 22 yr (profiles
%%on the left) and 220 yr (profiles to the right). \label{fig: pressure}}
%%\end{figure*}
% 

Direct radiation forces in astrophysical flows can strongly influence their evolution in scenarios like accretion 
onto compact objects or line-driven winds from massive stars.  In order to assess the radiation pressures exerted 
by Pop III stars upon their parent halos we added an operator-split update term to the ZEUS-MP momentum equation:  
\vspace{0.1in}
\begin{equation}
\frac{\partial}{\partial t}\,(\rho {\bf{v}}) = \frac{1}{c} \int_{\,\nu_{th}}^{\,\infty} \chi(\nu)\,{\bf{F}}(\nu)\,d\nu
\vspace{0.1in}
\end{equation} 
We utilized a flux corresponding to the blackbody spectrum of a 200 M$_{\odot}$ star with T$_{eff}$ = 10$^{5}$ K normalized 
to yield the emission rate of all photons above the hydrogen ionization threshold given in Table 4 of \citet{s02} and centered 
in the LCDM spherically-averaged halo discussed in section \ref{sect: uvesc}.
The neutral hydrogen component of the extinction coefficient exits the integral, which, along with the integral 
required to normalize the flux, only needs to be evaluated once. \raisebox{.4ex}{$\chi$} varies over ionization 
timescales because of its dependence on the neutral hydrogen fraction, so radiation momentum transfer into the gas 
must be accrued over consecutive chemical timesteps.  This is done by multiplying the force integral above in every 
problem zone by the current chemical timestep and summing this product over subsequent chemical times until the 
hydro timestep has been crossed.  The sum is then divided by the total density to obtain the update to the gas 
velocity in each zone.  This method neglects spectral hardening across the front that could only be captured by a
full multifrequency treatment of the radiation transfer, but these effects are likely to be unimportant in the high
densities of the primordial halo. 

Gas densities and UV fluxes are largest at the center of the halo so radiation forces exert their greatest influence in
the earliest stages of primordial H II region evolution.  UV photons can accelerate both the small neutral fractions  
in the postfront gas as well as the thin semi-neutral discontinuity at the front.  Gravitational, thermal pressure 
gradient, and radiation forces in the early I-front are compared at 22 yr and 220 yr in Fig \ref{fig: pressure}.  Thermal pressure 
gradients cancel gravity forces upstream of the shock leading the front (because of the hydrostatic initial 
conditions of the gas) but dwarf them, as expected, within the H II region itself.  The radiation force is also much 
greater than gravity within the H II region, an interesting result given that neutral fractions are less than 10$^{-5}$ there.  
Thermal pressure dominates radiation in the gas on either side of the I-front except in the partially-neutral layer of the 
front where the radiation force spikes because the product of the flux and neutral fraction peaks there.  In  
contrast, the radiation pressure is somewhat larger than thermal pressure in the postfront gas at 220 yr in this 
simulation, but their ratio in general is sensitive to changes in the very small neutral fraction in the ionized gas that can occur 
problem resolution or initial conditions.  Comparison of the direct UV forces upon the 
postfront gas at 22 yr and 220 yr clearly shows that they diminish with time.  As the 10$^{4}$ K gas 
expands, it is driven outward with an accompanying drop in central densities that suppresses recombination rates.  
Central neutral fractions fall (eventually to 10$^{-8}$), with a corresponding loss of radiation pressure.  Our
radiation force profiles are in general agreement with \citet{ket04}, who considered the same 200 M$_{\odot}$ 
blackbody spectrum but applied a different initial density profile.

Our simulations reveal that the radiation forces in the ionized gas behind the front enhance its velocity by
less than one percent, even though they are comparable to the thermal forces there.  This is due to the fact that 
the thermal pressure gradients in the shocked gas and just behind the front are primarily responsible for the 
acceleration of the flow, and these gradient forces are large in comparison to the radiation forces except 
across the few photon mean free paths (mfp) of the front.  Direct calculation of the UV acceleration of fluid elements
in the I-front layer itself is currently impractical for two reasons.  The mfp of UV photons through the baryonic 
densities typical of 10$^{6}$ M$_{\odot}$ halos is approximately 10$^{-7}$ pc, well below the resolution limit of 
an eulerian calculation that must accommodate the much larger dynamical scales of the H II region.  Failure to 
resolve the front can cause a code to interpret the very large radiation force peak to act upon the entire problem 
cell over the time required to ionize all of the cell, when in reality the force spike only operates on the fluid 
parcels in the extremely thin front for just the time required to ionize this layer (such an error led to 
unphysically large gas velocities in early trials of our code).  Furthermore, the intense UV flux in such proximity 
to the central star can violate the static approximation to the transfer equation, necessitating the use of the 
fully time-dependent equation to ensure accurate transport of the I-front, even across an appropriately rezoned grid. 

\begin{figure}
\resizebox{3.45in}{!}{\includegraphics{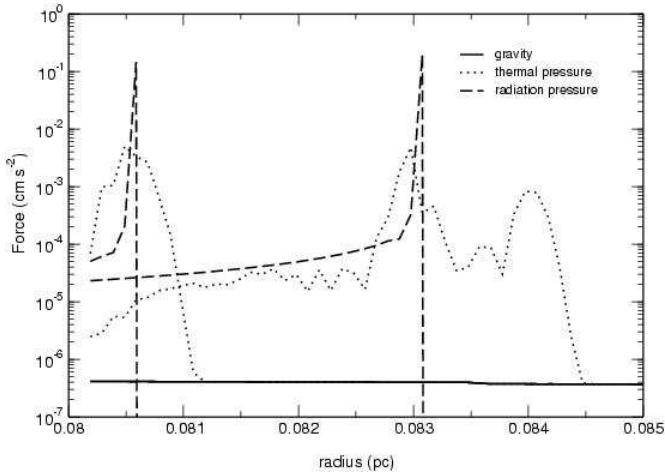}}
\caption{Thermal pressure gradient, radiation pressure, and gravitational accelerations taken at 22 yr (profiles
on the left) and 220 yr (profiles to the right).} \vspace{0.15in}
\label{fig:  pressure}
\end{figure}

A simple estimate of the radiative acceleration of gas elements in the front is possible by recognizing that it acts
on the layer for at most a few ionization timescales, between 10$^{4}$ s and 10$^{5}$ s at the times indicated in Fig 
 \ref{fig: pressure}.  The product of the radiation force peak in Fig \ref{fig: pressure} and these photoionization times yields an upper limit to the 
velocity imparted to the front, which we find from the data above to be less than 2 \kms, too little to significantly
alter the evolutionary outcome of the ionized flow.  This estimate is an upper bound because the force actually 
decreases as the parcels are ionized, while this procedure takes it to be constant (and at its maximum strength).  
When the thin layer is ionized, the large radiation force upon it evaporates and continues on to the next layer.  The
global radiation momentum transfer from the thin front into the gas accelerates it by at most 1 - 2 \kms.

\section{Improved Pop III UV Escape Fraction Estimates}  \label{sect: uvesc}
\begin{figure*}
\epsscale{1.08}
\plottwo{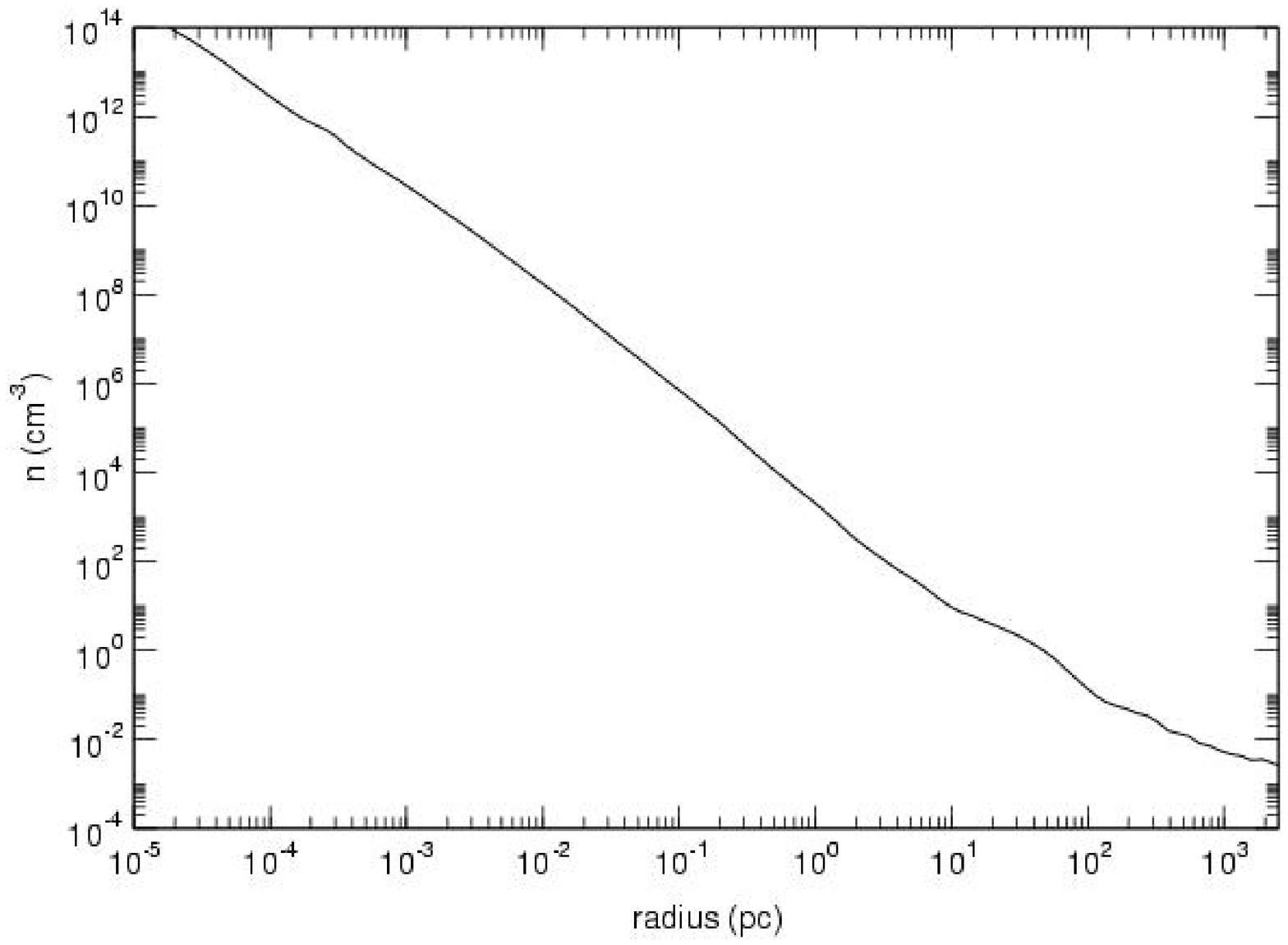}{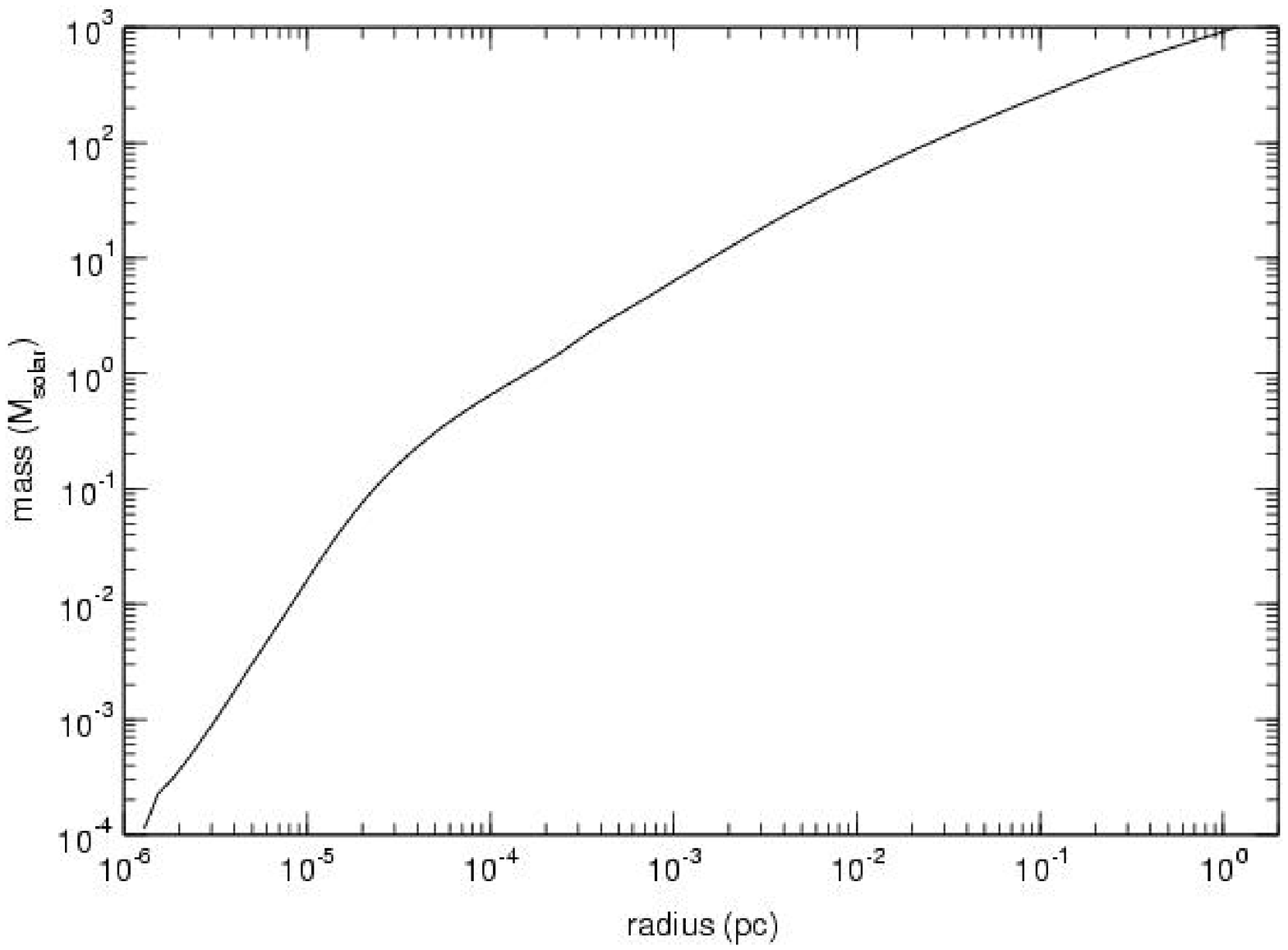}\vspace{0.15in}
\caption{Left panel: new spherically-averaged LCDM Pop III protostellar density profile \citep{oet05}.  
Right panel: baryonic mass enclosed by the given radii, corresponding to the profile on the left. \label{fig: dens/mtot}} \vspace{0.1in}
\end{figure*}

The breakout of UV radiation from Pop III minihalos was explored in an earlier paper \citep{wan04} (hereafter WAN04); 
we now extend this study to lower stellar masses with improved problem setup.  We adopted the spherically-averaged
LCDM baryon field of a 5 $\times$ 10$^{5}$ M$_\odot$ dark matter minihalo formed at 
z = 18 for our initial number density profile \citep{oet05} in place of the CDM density profile of \citet{abn02}.  Also, in our 
earlier work, the inner boundary was set to 0.06 pc in the CDM protostellar density, regardless of the mass assumed 
for the central star later forming at its center.  Launching the I-front from this radius would neglect the 100 - 
200 M$_\odot$ gas mass interior to this radius not consolidated into the protostar in the lower stellar mass cases, 
as seen in Fig \ref{fig: dens/mtot}.  The inner boundary was therefore set to be the radius enclosing a gas mass equal 
to the mass 
of the star whose UV escape fraction was being studied.  Although this inner few hundred M$_\odot$ is small in comparison 
to the outer tens of thousands of solar masses later ionized in our previous studies, they were included because 
their significantly higher number densities and recombination rates might disproportionately retard the advance of 
the front.  Apart from these two modifications, the problem setup was the same as in the WAN04 simulations.  

An important consequence of shifting the inner problem boundary in accord with the protostellar mass is that the 
nascent I-front can now encounter much higher central densities, up to 10$^8$ cm$^{-3}$ for the 25 M$_\odot$ runs 
compared to the 10$^4$ cm$^{-3}$ densities in WAN04.  The accompanying rise in photoionization rates restricts the 
timestep control to much shorter advances, making 25 M$_\odot$ the lower practical computational limit in these 
density profiles.  Despite gathering interest in the physical processes that cut off accretion onto Pop III
protostars \citep{oi02,op03,ra04}, they are not understood well enough to provide firm estimates on the final masses 
of these objects, only that they are likely 30 - 300 M$_\odot$ \citep{abn02,bcl99}.  Future models will  
also be needed to determine the central envelope densities and infall through which primordial I-fronts truly emerge; 
the exit of ionization fronts through accretion infall will be studied in forthcoming 3-D simulations.

\begin{deluxetable}{ccccc}
\tabletypesize{\scriptsize}
\tablecaption{Final I-front radii, breakout times, and f$_{esc}$.\label{tbl-3}}
\tablehead{
\colhead{M$_{\star}$ (M$_\odot$)} & \colhead{radius (pc)} 
                    & \colhead{t$_{brkout}$ (yr)} & \colhead{r$_{brkout}$ (pc)} & \colhead{f$_{esc}$}}
\startdata 
  25   &  5.0e-3  &  ---   &  ---  &    0      \\
  40   &  8.8e-3  &  ---   &  ---  &    0      \\
  80   &  1978  &  3.3e05  &  7.9  &   89\%   \\
  120  &  2633  &  2.3e05  &  7.2  &   91\%   \\
  140  &  3186  &  1.7e05  &  4.9  &   93\%   \\
  200  &  3505  &  1.3e05  &  3.7  &   94\%   \\
  260  &  3856  &  9.0e04  &  2.8  &   96\%   \\
  300  &  4441  &  8.5e04  &  2.6  &   96\%   \\
  400  &  4665  &  5.9e04  &  2.3  &   97\%   \\
  500  &  5132  &  1.9e04  &  1.0  &   99\%   \\
\enddata
\end{deluxetable}

Results for stellar masses ranging from 25 M$_\odot$ to 500 M$_\odot$ using the time-averaged photon rates from Table 
4 of \citet{s02} appear in Table 3.  Since no UV photons exit the
virial radius of the halo before the front transforms from D-type back to R-type, f$_{esc}$ is simply
\vspace{0.05in}
\begin{equation}
 f_{esc} = \frac{t_{msl} - t_{breakout}}{t_{msl}}\vspace{0.05in}
\end{equation}
where t$_{msl}$ and t$_{breakout}$ are the stellar 
main sequence lifetime and time to I-front breakout, respectively.  Comparison of these final I-front radii with those
of WAN04 for M $>$ 100 M$_\odot$ indicates that they differ by 20\% at 120 M$_\odot$ but only by 4\% at 500 M$_\odot$ 
in a trend toward greater agreement with mass.  WAN04 excluded the most intervening mass at 120 M$_\odot$ and the least 
at 500 M$_\odot$, so it is not surprising the I-front is delayed the most in the first case.  As expected, breakout 
radii and times decrease with increasing stellar mass

The I-front breakout from the 80 M$_\odot$ case shows that lower-mass Pop III stars still effectively ionize their local
environments and can establish ionized outflows capable of dispersing any elements ejected by pulsational instabilities
before the star collapses directly into a black hole \citep{hw03}.  In contrast, the primordial envelope traps the 
I-fronts of the 25 M$_\odot$ and 40 M$_\odot$ stars to form ultracompact H II (UC H II) regions with lifetimes of 6.5 Myr
and 3.9 Myr, respectively.  The higher central recombination rates overwhelm the lower ionizing photon rates of a lower 
mass star to stall the front at subparsec radii.  A star in this mass range cannot remain on the main sequence long enough 
for the ionized pressure bubble to expand and free the front.

\section{Conclusion}

We have presented an explicit multi-step scheme for integrating the coupled equations of radiation transport,
ionization kinetics, radiative heating and cooling, and hydrodynamics.  We have validated our method against
a large battery of analytic test problems for both static and moving media.  While these are 1D tests, the 
method is easily extended to 3D.
An important strength of our multistepping algorithm is its applicability to a variety of radiation transport schemes for 
computing photoionization rates on a grid.  Many problems can be addressed by this method in its current state as 
a single source code, with more soon being possible with the planned activation of the full 9-species network 
as well as multifrequency and multisource upgrades.  Full 3--D radiation hydrodynamical trials of the code are 
currently underway to simulate the escape of UV radiation from high-redshift minihalos formed from realistic
cosmological initial conditions.  One important question to be answered is whether the UV escape fraction cutoff 
observed by Kitayama, \etal with low stellar masses persists if three-dimensional instabilities arise at the 
ionization front.  Instabilities may open channels out of the halo and permit the exit of UV radiation that 
would not otherwise have escaped, enhancing UV escape fractions from low mass stars.  On the other hand, recent 
high dynamical range AMR simulations of primordial star formation \citep{oet05} have 
revealed the formation of disks around protostars.  These disks may reduce escape fractions even from high mass 
stars by blocking UV photons along lines of sight in the plane of the disk with large optical depths.  The radiation 
hydrodynamical evolution of these objects will enable the simulation of more realistic initial conditions for the 
explosion of supernovae and subsequent chemical enrichment of the early universe.  Such simulations would also set
the stage for self-consistent modeling of the energetics of cosmological miniquasars and the propagation of their
radiation into the early IGM \citep{km05}. 

While we have modified our algorithm to study the photoevaporation of minihalos in the vicinity of a massive 
Population III (Pop III) star in 1--D at high redshifts \citep{oet05}, the ionization of these objects in 3--D with 
primordial chemistry remains to be done.  The catalysis of molecular hydrogen within these structures in the presence 
of SUV and x-ray backgrounds as they are being ionized by external fronts is a key issue in radiative feedback 
processes in the early universe that is not well understood.  Another process within the realm of study of our single
source algorithm is the potential cutoff of accretion onto Pop III protostars by nascent I-fronts as the star enters
the main sequence.  Radiation hydrodynamical simulation of accretion cutoff processes with improved stellar evolution
models at their foundation will provide firmer estimates of the true mass spectrum of Pop III stars, a key ingedient
in large scale calculations of early reionization.  Studies of the escape of UV radiation from high redshift 
protogalaxies of $\sim$ 1000 Pop III stars are currently being planned in connection with the addition of cartesian
multisource VTEF radiative transfer \citep{pn05} to ZEUS-MP.  Having exhaustively validated our coupling scheme for 
radiative transfer and hydrodynamics in 1--D with the array of static and hydrodynamic tests presented in this paper,
we can now apply the algorithm to these relevant and timely problems in three dimensions with confidence.

\acknowledgments

We would like to thank the anonymous referee for suggestions which have significantly improved the quality of this 
paper.  DW would like to thank Mordecai Mac-Low, Tetsu Kitayama, and Ilian Iliev for useful discussions.  This work 
was supported in part by NSF grants AST-0307690 and AST-9803137.  DW has also been funded in part by the U.S.\ Dept.
\ of Energy through its contract W-7405-ENG-36 with Los Alamos National Laboratory.  The simulations were performed 
at SDSC and NCSA under NRAC allocation MCA98N020.

\end{document}